# High-Speed Boulders and the Debris Field in DART Ejecta

Tony L. Farnham[1,2], Jessica M. Sunshine[2,3], Masatoshi Hirabayashi[4], Carolyn M. Ernst[5], R. Terik Daly[5], Harrison F. Agrusa[6], Olivier S. Barnouin[5], Jian-Yang Li[7], Kathryn M. Kumamoto[8], Megan Bruck Syal[8], Sean E. Wiggins[8], Evan Bjonnes[8], Angela M. Stickle[5], Sabina D. Raducan[9], Andrew F. Cheng[5], David A. Glenar[10], Ramin Lolachi[11,10,12], Timothy J. Stubbs[10], Eugene G. Fahnstock[13], Marilena Amoroso[14], Ivano Bertini[15], John R. Brucato[16], Andrea Capannolo[17], Gabriele Cremonese[18], Massimo Dall'Ora[19], Vincenzo Della Corte[20], J. D. P. Deshapriya[21], Elisabetta Dotto[21], Igor Gai[22], Pedro H. Hasselmann[21], Simone Ieva[21], Gabriele Impresario[14], Stavro L. Ivanovski[23], Michèle Lavagna[17], Alice Lucchetti[18], Francesco Marzari[18], Elena Mazzotta Epifani[21], Dario Modenini[22], Maurizio Pajola[18], Pasquale Palumbo[15], Simone Pirrotta[14], Giovanni Poggiali[16], Alessandro Rossi[24], Paolo Tortora[22], Marco Zannoni[22], Giovanni Zanotti[17], Angelo Zinzi[25]

[1] Corresponding author farnham@umd.edu.
[2] *University of Maryland, Department of Astronomy, College Park, MD 20742, USA.*
[3] *University of Maryland, Department of Geology, College Park, MD 20742, USA.*
[4] *Daniel Guggenheim School of Aerospace Engineering, Georgia Institute of Technology, 620 Cherry Street, Atlanta, GA 30332, USA.*
[5] *Johns Hopkins University Applied Physics Laboratory, Laurel, MD 20723, USA.*
[6] *Université Côte d'Azur, Observatoire de la Côte d'Azur, CNRS, Laboratoire Lagrange, Nice, France.*
[7] *Planetary Science Institute, 1700 East Fort Lowell, Suite 106, Tucson, AZ 85719, USA.*
[8] *Lawrence Livermore National Laboratory, Livermore, CA 94550, USA.*
[9] *Space Research and Planetary Sciences, Physics Institute, University of Bern, Bern, Switzerland.*
[10] *NASA Goddard Space Flight Center, Greenbelt, MD 20771, USA.*
[11] *Center for Space Sciences and Technology, University of Maryland, Baltimore County, Baltimore, MD 21250, USA*
[12] *Center for Research and Exploration in Space Science and Technology, NASA/GSFC, Greenbelt, MD 20771, USA*
[13] *Jet Propulsion Laboratory, California Institute of Technology, Pasadena, CA 91109, USA.*
[14] *Agenzia Spaziale Italiana (ASI), Via del Politecnico snc, 00133, Rome, Italy*
[15] *Department of Science and Technology, University of Naples 'Parthenope', Naples, Italy.*

[16] *INAF-Osservatorio Astrofisico di Arcetri, Firenze, Italy.*
[17] *Politecnico di Milano, Dipartimento di Scienze e Tecnologie Aerospaziali, Milano, Italy.*
[18] *INAF-Osservatorio Astronomico di Padova, Padova, Italy.*
[19] *INAF-Osservatorio Astronomico di Capodimonte, Napoli, Italy*
[20] *INAF-Istituto di Astrofisica e Planetologia Spaziali, Roma, Italy.*
[21] *INAF-Osservatorio Astronomico di Roma, Via Frascati 33, Monte Porzio Catone, RM, Italia.*
[22] *Università di Bologna, Bologna, Italy*
[23] *INAF-Osservatorio Astronomico di Trieste, Trieste, Italy.*
[24] *IFAC- Istituto di Fisica Applicata Nello Carrara, Sesto Fiorentino, Italy.*
[25] *ASI Space Science Data Center (ASI-SSDC), Via del Politecnico snc, 00133, Rome, Italy*

Abstract

On 26 September 2022 the Double Asteroid Redirection Test spacecraft (DART) collided with Dimorphos, the moon of the near-Earth asteroid 65803 Didymos, in a full-scale demonstration of a kinetic impactor concept. The companion LICIACube spacecraft documented the aftermath, capturing images of the expansion and evolution of the ejecta from 29 to 243 s after the impact. We present results from our analyses of these observations, including an improved reduction of the data and new absolute calibration, an updated LICIACube trajectory, and a detailed description of the events and phenomena that were recorded throughout the flyby. One notable aspect of the ejecta was the existence of clusters of boulders, up to 3.6 m in radius, that were ejected at speeds up to 52 m s$^{-1}$. Our analysis of the spatial distribution of 104 of these boulders suggests that they are likely the remnants of larger boulders shattered by the DART spacecraft in the first stages of the impact. The amount of momentum contained in these boulders is more than 3 times that of the DART spacecraft, and it is directed primarily to the south, almost perpendicular to the DART trajectory. Recoil of Dimorphos from the ejection of these boulders has the potential to change its orbital plane by up to a degree and to impart a non-principal axis component to its rotation state. Damping timescales for these phenomena are such that the Hera spacecraft, arriving at the system in 2026, should be able to measure these effects.

1. Introduction

NASA's Double Asteroid Redirection Test (DART) was devised as an experiment to demonstrate the use of a kinetic impactor to deflect the trajectory of a potentially hazardous object that might threaten the Earth. The primary goal was to measure the momentum enhancement, beyond that of the spacecraft itself, that is produced by asteroidal material ejected in the collision (Rivkin et al. 2021). On 26 September 2022, the DART spacecraft impacted asteroid Dimorphos, the secondary component of the binary asteroid system (65803) Didymos, changing its orbital period by 33±1 (1σ) minutes (Daly et al. 2023; Thomas et al. 2023). The successful outcome of the experiment proved the feasibility of the kinetic impactor concept, and found a momentum enhancement in the direction of Dimorphos' orbital velocity vector of $\beta = 3.6^{+0.19}_{-0.25}$ (Cheng et al. 2023) but this result is only the beginning of our understanding of the full-scale kinetic impact deflection. There has yet to be a full accounting of the total momentum in all directions, but the ejecta cone spreads out sideways as well as in the direction of the incoming spacecraft. Thus, a significant component of the momentum, possibly several times that contributing to the β factor, was carried out perpendicular to Dimorphos' velocity vector. To fully understand the impact's effect on Dimorphos' orbit, it is necessary to explore the spatial distribution of the debris field and use it to ascertain the net momentum of all its components.

DART's companion spacecraft, LICIACube (Light Italian Cubesat for Imaging of Asteroids) (Dotto et al. 2021), separated from DART 15 days before the impact. A few minutes after the impact, it flew past Didymos and recorded the aftermath of the event (Dotto et al. 2024). The images obtained reveal a generally conical structure made up of small (e.g., unresolved) dust grains that form intricate and complex filaments streaming away from the asteroid. In addition to the filamentary structures, LICIACube also detected clusters of distinct point source features that can be tracked through as many as 21 different images. These point sources appear to be boulders ejected by the impact and could represent a notable fraction of the ejecta mass. Determining the speeds and directions in which they were emitted will provide an understanding of the mechanisms that controlled the impact physics, and will provide additional constraints on the momentum imparted from the impact.

Previous studies (Cheng et al. 2023; Deshapriya et al. 2023; Hirabayashi et al. 2024) have modeled the dust ejecta as a geometrical unit to determine its general shape and orientation, finding that it is essentially conical, though the North-South (relative to Dimorphos) opening angle, at ~135°, is larger than its ~100° East-West opening. While showing relative consistency, the dust ejecta cone's central axis is slightly offset from both the incoming DART trajectory and the surface normal. Although these models describe the overall structure of the ejecta and are useful for investigating the bulk direction of the momentum vector, they are limited in what they reveal about the details of the ejecta itself. Further constraints on the positions and speeds of the debris and the boulder clusters, dust filaments, and clumps that deviate from the general cone structure, that will inform about the mechanisms at work during the impact process. Determining the locations and velocities of the features to map the detailed structure of the debris field in three dimensions will allow us to investigate how the ejecta evolved during the flyby. Fortunately, the LICIACube observations are an excellent dataset for obtaining the necessary 3-D measurements. Because images were recorded throughout the flyby, the changing viewpoint introduces parallax that can be used to accurately locate features in space. Furthermore, because the ejecta exhibits such complex and detailed structures, including numerous clumps, kinks and intersections, as well as individual boulders, there are many features that can be used to map the field, compute real velocities, and track its temporal evolution.

This paper addresses the three dimensional distribution of the ejecta cloud, how the morphology evolves over the first few minutes after impact, and how it contributes to the total momentum enhancement. We discuss the LICIACube observations and our improved data reduction technique and then offer a tour of the encounter, with chronological descriptions of the events and features that were recorded. Measurements of the boulders seen in the images are used to derive their locations and velocities, as well as their contributions to the momentum budget and the implications therein. We also provide the derivation of the parallax measurements in Appendix 1 and describe our calculation of a new LICIACube trajectory in Appendix 2. Ultimately, this work, and subsequent analyses of the ejecta cone, will provide a detailed map of the debris field that can be used to determine the properties of its constituent components (e.g., velocities, temporal changes in morphology) that will help to understand the impact physics and result in a more complete understanding of the kinetic impactor concept.

## 2. Observations and Data Reduction
### 2.1. LICIACube Data

The LUKE (LICIACube Unit Key Explorer) instrument is a wide-field (9.2° × 4.9°) imager with an RGGB Bayer pattern filter (Dotto et al. 2021). The LUKE dataset consists of 183 images obtained from 20 sec to 243 sec after the DART impact, with the closest approach (C/A) to Didymos at 57.6 km occurring at I+166.94 sec (see Appendix 2). Observations were generally obtained in triplets, where images with three different exposure times (short, intermediate and long) were captured in rapid succession, at intervals of 6 sec (early and late in the flyby), 3 sec (within ~30 sec of C/A), and 1 sec (within 12 sec of C/A). The exposure times within each triplet varied from one sequence to the next, so the images exhibit a wide range of signal-to-noise (S/N) as well as different levels of saturation. The LUKE camera uses a Bayer filter, in which clusters of four pixels record different colors (1 red, 2 green and 1 blue) and images are reconstructed for each color by interpolating over the pixels of the other colors. For the work presented here, we used only the red filter images, for two reasons: 1) Our analysis showed that they have the narrowest point spread function (PSF) and thus produce the sharpest features, and 2) the contrast of boulders and other details is better in the red than in the other colors.

The spacecraft attempted to track the asteroid throughout the flyby, and though largely successful, Dimorphos moved out of the field of view in the closest approach frames from I+161 to I+172 sec and Didymos was only partially in the frame, in the lower left corner of the images, from I+164 to I+169 sec. Thus, although the ejecta "above" Didymos in the images was captured throughout the encounter, the material "below" the asteroid is not visible around close approach, which affects our ability to map the ejecta in those regions.

The LEIA (LICIACube Explorer Imaging for Asteroid) instrument (Dotto et al. 2021), a narrow-field camera, obtained 210 images during the flyby (Zinzi & Della Corte 2023a). Unfortunately, due to an undiagnosed problem that occurred during flight, the observations are out of focus and thus of poorer quality than the LUKE images, and are not used in this work.

## 2.2. Data Reduction and Image Merging

Although calibrated images are available from the Planetary Data System (PDS), the bias/dark removal (which uses pre-launch data) leaves residual vertical striping and large regions where the background is truncated at zero flux. Because these artifacts interfere with our ability to identify and track the intricate features and boulders, we opted to reduce the data ourselves to improve the image quality. We started with the raw data from the PDS (Zinzi & Della Corte 2023b) and used similar reduction procedures as outlined in the PDS calibration document (Zinzi & Della Corte 2023c) with two significant exceptions. First, rather than using the ground-based bias and dark calibration frames, we used a short exposure image of blank space (frame 1664234361.0008), obtained about 36 seconds after the last Didymos image, to represent the bias and dark levels.

Although this image may have a lower dark signal and a slight additional signal from the sky background, our tests showed that these issues introduced no noticeable problems, and in fact, because the frame was obtained at nearly the same time and temperature as the asteroid data, it produces a much better background removal than the ground-based calibration frames scaled for the detector temperature (Figure 1). The improvements facilitate our ability to track boulders over multiple images. The second difference in the reduction is that we derived our own absolute calibration coefficients directly from observations that LUKE obtained on 22 September 2022 of the secondary spectrophotometric standard star Xi$^2$ Ceti. Using the camera and optical characteristics provided in the LUKE documentation (Zinzi & Della Corte 2023c), and the spectrophotometric calibrations from Hamuy et al. (1992), we found radiance conversion coefficients for the three color filters, in units of (W m$^{-2}$ nm$^{-1}$ sr$^{-1}$)/(DN/s): Red, $8.0\times10^{-8} \pm 1.2\times10^{-8}$; Green, $9.1\times10^{-8} \pm 1.7\times10^{-8}$; and Blue, $1.3\times10^{-7} \pm 6.9\times10^{-9}$. As a check, we confirmed that the radiance levels derived from our calibrations are consistent with those derived by Lolachi et al. (2025) (who independently used observations of the Pleiades to compute a correction to the calibration used in the PDS images), and also compared the derived surface radiance of common spots on Didymos to the radiance levels measured by the DART spacecraft, with good agreement.

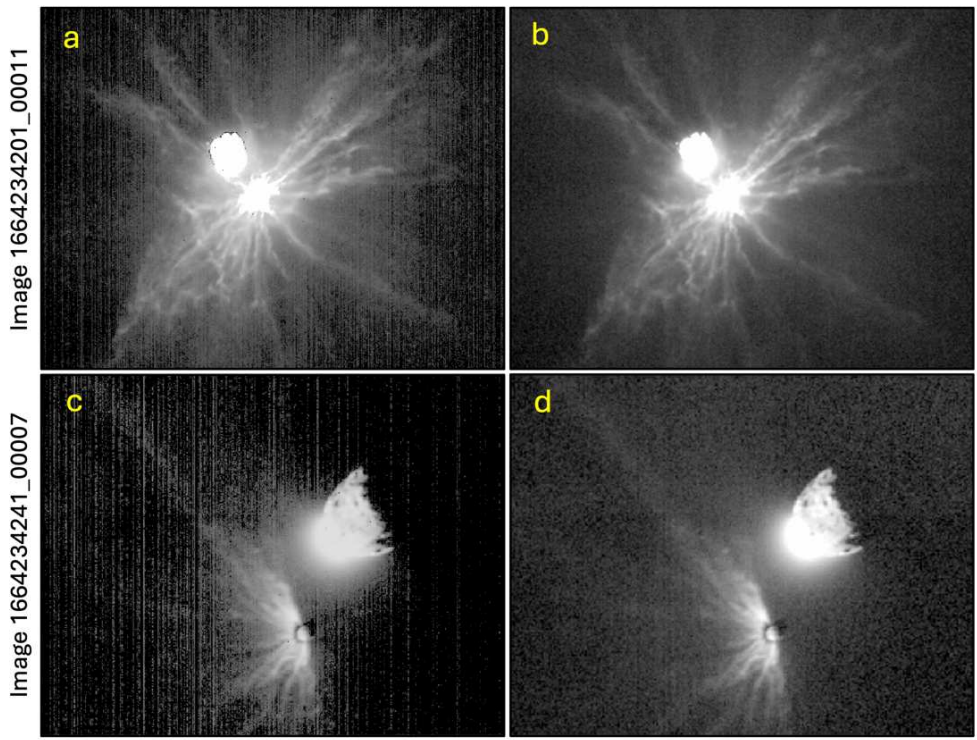

Figure 1. Comparison of calibrated images available in the PDS (a & c) to the new reductions used in this work (b & d). The new version removes most of the artificial structure in the background, improving the visibility of details in the ejecta morphology. The common images are shown with the same display levels and contrast stretch.

In order to maximize the utility of the images, we needed to adopt the best S/N portions of the data, while minimizing or eliminating the saturated regions. To this end, we merged the data from each triplet of exposures to produce an optimum set of images. The three triplet frames were obtained within a fraction of a second, so the viewing geometry is essentially the same, with the exception for those at closest approach, where the geometry between the first and last images changed by at most a degree (and Dimorphos and most of the ejecta were outside the field of view). For the purposes of this study, these small changes are negligible. We first registered the three frames in each triplet, either aligning features on Didymos that are common between two frames, or when Didymos was saturated, using ejecta features close to Didymos for the alignment. Next, we merged the three images into a single frame: Starting with the longest exposure to capture the faintest features, we then replaced any saturated or near-saturated pixels with unsaturated pixels from the intermediate exposure image. Any remaining saturated pixels were then replaced by those from the short exposure to produce the merged frame. There are a few cases where all three images in the triplet are not available, so a few of the merged frames still have saturated regions (for a missing short exposure), while some have low S/N (where the long exposure is missing). Because the frames were first converted to radiance, the brightness levels are consistent, and in general, the images merged well. Where pixel segments were replaced, the seams are sometimes discernible, but other than along the sharp limb of Didymos in a few images, they are not prominent. For the acquisition time of the image, we adopted the time of the longest exposure that was included, because most (and in some cases, all) of the pixels come from that frame. These merged images were used for the work presented here, and we remain cognizant of the possibility that the seams have the potential to introduce artifacts that may be interpreted as features. Thus, whenever odd or unusual features were encountered, we returned to the original, individual images to confirm any interpretations. Both the newly calibrated images and the merged images are archived in the PDS (Farnham 2025a).

To provide additional insight into the temporal aspects of the flyby, we created animated sequences from the merged images. A number of these movies were generated, with different parameters that change the focus and/or emphasis of the result (e.g., centered on Didymos, centered on Dimorphos, scaling to simulate a fixed spacecraft range, etc.). These movies are valuable tools for the evaluation and interpretation of the data and how it changes with time (due to physical evolution as well as the changing viewpoint), but any measurements that are obtained come from the individual images. Each frame of the movie is tagged with the time since impact[1], to provide a reference that can be used to identify the events discussed in the following analyses. These animated sequences are available in MOV and MP4 format and can be accessed from the University of Maryland Digital Repository at https://doi.org/10.13016/ut5x-zjtz (Farnham 2025b). They can be used to supplement the figures shown below to illustrate the phenomena that are observed throughout the encounter.

### 3. General Overview of the Flyby

The trajectory of the LICIACube spacecraft was essentially parallel to that of the DART spacecraft, but offset by 57.3 km and delayed by 167 sec (See Appendix 2). Due to the offset, LICIACube passed over the southern hemisphere of Didymos (whose North pole is oriented south of the ecliptic), reaching an extreme sub-spacecraft latitude of -59°. The viewpoint at close approach was almost the same as those from Earth, HST and JWST (e.g., Li et al. 2023; Thomas et al. 2023), allowing good comparisons of the different observations, but at drastically different resolutions. In this section, we present a chronological discussion of the events that were recorded in the LUKE images and introduce some of the interesting phenomena that were observed. Figure 2 is an animated sequence of our merged images and illustrates the phenomena discussed. We use terminology such as "above" or "below" to refer to directions in the images under discussion, because of the confusion that can arise in body-fixed directions due to the retrograde spin of Didymos and the constantly changing viewpoint. (For reference, Didymos' South pole, projected onto the sky, is oriented toward the upper right during the approach phase, and rotates around the top to the upper left during the departure.) The time since impact provides the temporal reference used in the discussions.

---

[1] To determine the LUKE image number based on the time since impact: Triplet image number = 1664234064 + round(time since impact). The three frames of different exposures in that triplet are identified by additional extensions in the filename.

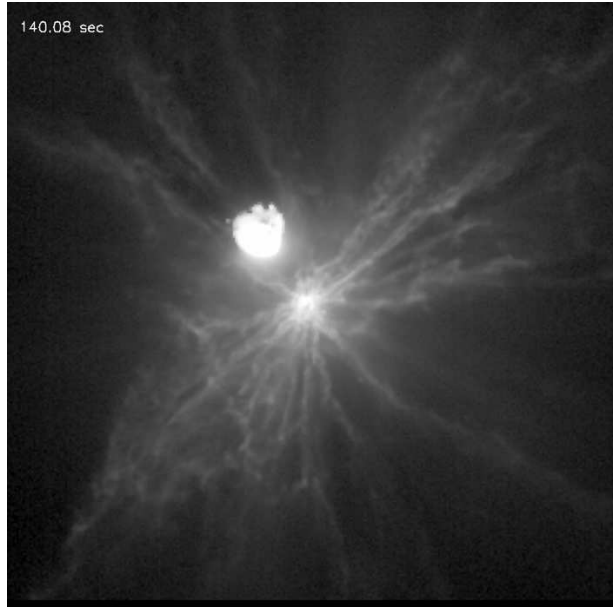

Figure 2. Animated sequence of LUKE images showing the evolution of the ejecta morphology during the flyby. The sequence covers the period from 29 sec to 243 sec after impact (times are shown in the upper left corner), with the initial face-on dust distribution transforming into a wedge shape as the spacecraft moves from inside to outside the dust cone around close approach. Images are centered on Dimorphos and have been scaled to a common spacecraft range, with a field of view spanning 7.80 km. The black sections of the images near closest approach are regions that lie beyond the edges of the detector.
[This figure links to an animated sequence.]

### 3.1. Approach Phase

We define the approach phase of the encounter from 29 to 140 sec, when LICIACube's trajectory was primarily radial with respect to Didymos. During this time, the spacecraft range decreased dramatically, but there was little change in the viewpoint. In this phase, the spacecraft remains situated inside the projected edges of the dust cone. The filaments and streamers rise outward and toward the observer, with Dimorphos lying behind the nexus and Didymos in the background (Figure 3). As the range decreases (see Figure 2), features can be detected on Didymos, including a jagged terminator (due to rough terrain near the pole) and a concave segment along the limb that indicate that the asteroid is not strictly triaxial in shape. There is also a bright spot beyond the terminator to the upper left that shows there is a topographic high point that extends into sunlight. Because the viewpoint is nearly constant, changes detected in the ejecta are dominated by the outflowing motions of the debris. Thus, these observations are ideal for directly measuring the projected velocities of the early ejecta. From 29 to 50 sec, the camera's field of view extended well beyond the ejecta cone, so even the fastest clumps remain within the field, but the low resolution at this time means that only large blobs are detected. Using the motions of the features in the first four images (from 29 to 47 sec) we measured the speeds of two of the fastest clumps (Figure 3). Feature #1 is the fastest identifiable clump, with a speed 340 m s$^{-1}$, while #2 moves at 240 m s$^{-1}$. However, because they are projected onto the sky plane, these are lower limits to the actual speeds. They agree well with measurements of the same features (labeled S1 and S2) by Dotto et al. (2024).

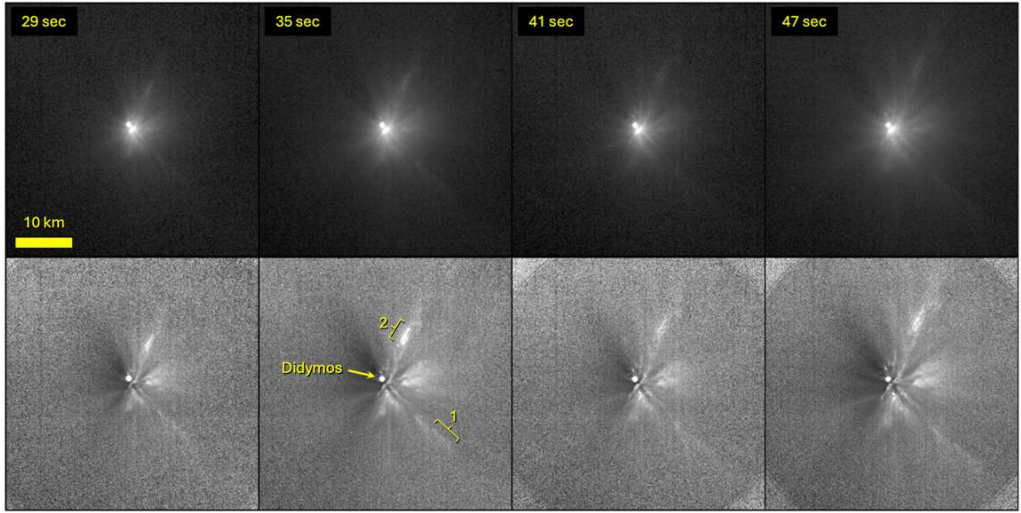

Figure 3. The first four LUKE images, scaled to the same spacecraft range, showing the approach morphology and the large, rapidly expanding structures in the early ejecta. The upper panels show the images with a log display, while the lower panels have been enhanced by dividing out an azimuthally averaged radial profile to reveal the details in the dust cloud. The highlighted features have projected speeds of 340 (#1) and 240 (#2) m s$^{-1}$.

      As the range decreases, the fastest debris exits the field of view, while at the same time, the improving resolution reveals highly detailed features comprising the filaments (e.g., Figure 2). These convoluted structures were produced by the interplay between the surface topography interfering with the outflowing material and the effects that the outflowing material have on the

surface during the excavation of the impact crater (e.g., as the crater develops, the topography changes, which in turn changes the directions of ejecta outflow (DeCoster et al. 2024; Raducan et al. 2024)). Evidence of a similar filamentary outflow is seen at smaller scale on Dimorphos' pre-impact surface (Sunshine et al. 2024). The streamers seem to be more extensive and well-defined in Dimorphos' North–South quadrants (upper right–lower left) than in the East–West quadrants. It is likely that this phenomenon is related to the dichotomy in the N–S and E–W opening angles found in the dust cone studies (Deshapriya et al. 2023; Hirabayashi et al. 2024), and is probably related to Dimorphos' smaller radius of curvature in the N–S direction (Daly et al. 2024). See Section 3.2 for additional discussion.

## 3.2. Flyby Phase

We define the flyby phase of the encounter from 140 to 195 sec, when LICIACube's trajectory was primarily translational, with little change in the range but dramatic changes in viewing geometry. This phase straddles closest approach, so it provides not only the highest resolution images, but also the best parallax between images. The numerous kinks and clumps that are visible in the ejecta provide excellent reference points for tracking these features and mapping the debris field. Thus, the data from this phase are the primary focus of our 3-D analyses.

The general progression of events during this phase starts with the spacecraft inside the dust cone. However, unlike the approach phase, the translational motion reveals the dimensionality of the scene as the streamers move in front of Didymos (Figure 4a-c). In the pre-C/A images of this phase, the limb of Dimorphos can be detected peeking out from a gap in the streamers to the lower right (Figure 5). Furthermore, there appears to be a shadow on Dimorphos' disk that is being cast by the nearby ejecta (compare to Didymos' disk in Figure 4c for the illumination conditions that would be expected with no ejecta). This shadow provides the first evidence that portions of the ejecta cloud are optically thick, even though regions between the streamers are nearly transparent.

In the pre-C/A images, more than 100 boulders are detected and tracked from one image to the next, appearing in as many as 21 different frames from 131 to 168 sec. Interestingly, these boulders are not uniformly distributed (Figure 6). The majority are located in two clusters: a large one in the upper right quadrant and a smaller one to the lower left. There are also a few individual boulders scattered throughout the lower right quadrant. The clustering suggests that these objects were ejected in preferred directions. Additional discussion of these boulders, including a 3-D analysis of their spatial distribution and a description of the pertinent observational selection effects, is presented in Sections 4 and 5.

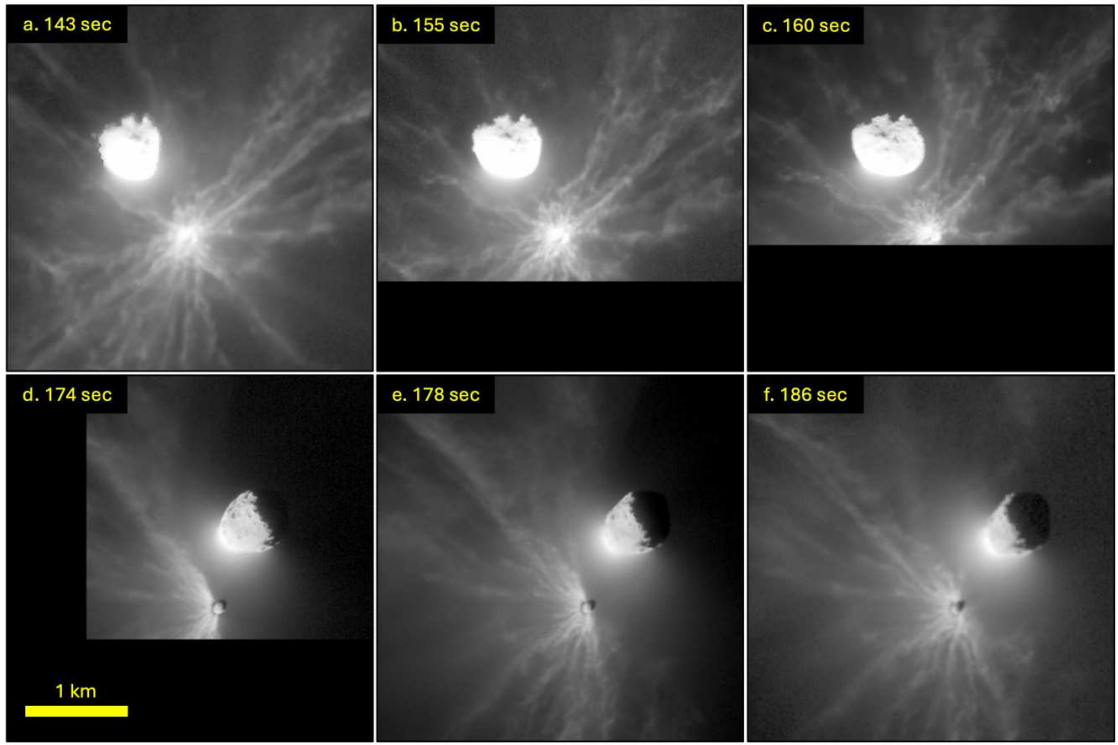

Figure 4. Sequence of merged images showing the ejecta during the flyby phase. Panels a-c show pre-C/A views, with LICIACube inside the ejecta cone. Filaments can be seen moving in front of Didymos as the viewpoint changes. Panels d-f show the post-C/A views, after the spacecraft moved outside the cone. In panel d the viewpoint is nearly perpendicular to the cone axis; in e the viewpoint extends along the edge of the cone; and in f the cone is visible entirely surrounding Dimorphos. Images have been scaled to a common spacecraft range. The black regions are the sections of the image outside the edges of the detector.

From 161 to 172 sec, Dimorphos is completely outside the field of view, while from 164 to 169 sec, portions of Didymos are also absent (though Didymos never completely leaves the field). Unfortunately, around these times, most of the ejecta cloud is also out of the field of view and thus, information from the time of close approach is missing. It is during this gap in coverage that LICIACube moves from inside the ejecta cone to outside, and when Dimorphos reappears at 173 sec, the debris field exhibits a very different appearance.

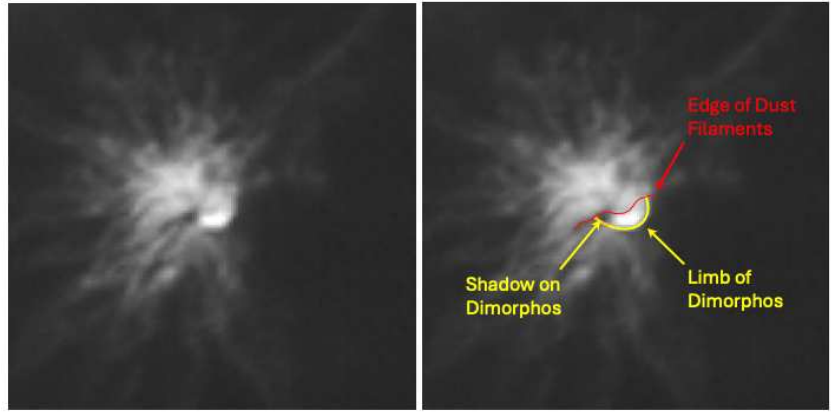

Figure 5. Close-up of the ejecta cloud at 159 sec with Dimorphos visible through a gap in the filaments. The limb and shadow are highlighted. In the absence of ejecta, the portions of Dimorphos that are visible here would be fully illuminated (e.g., bright limb; compare the disk of Didymos in Figure 4c).

From 173 to 175 sec, a classic ejecta cone viewed nearly side-on can be seen, with bright wings spanning an apparent angle of ~150° and streamers filling in the region in between (Figure 4d). It is interesting to note that there are some streamers several kilometers away that lie outside this cone, emphasizing the enigmatic nature of the debris field and hinting at the complexities involved in the impact processes (e.g., Barnouin et al. 2019). Deshapriya et al. (2023) and Hirabayashi et al. (2024) projected the wings of this ejecta cone linearly back to Dimorphos, where they converged at a point near the center of the body, suggesting that the impact produced a very deep crater. However, in the image from 174 sec (nearest C/A, when the cone is close to perpendicular to the line of sight), the wings appear to bend inward as they converge on Dimorphos, (Figure 7a & b) which is an indication that the ejecta cone widens with time. This could be a natural decrease in the ejecta angle as the crater expands (e.g., Anderson et al. 2003) or it could be explained by the small size of Dimorphos, whose surface curves away from the impact site. If the ejecta angle, $\alpha$, between the excavation front and the local surface remains constant with time as the crater grows, then the angle between the excavation front and the cone axis, $\theta$, must increase (Figure 7) causing the cone to widen with time. When the apparent bending of the wings is taken into account, the cone converges at a shallower depth than the linear projection would imply, discounting the need for a deep crater (Hirabayashi et al. 2024). This same dependence on the surface curvature also explains the difference in the cone's widths in the N-S and E-W directions, as noted above, because Dimorphos' radius is smaller in the N-S direction than along its equator.

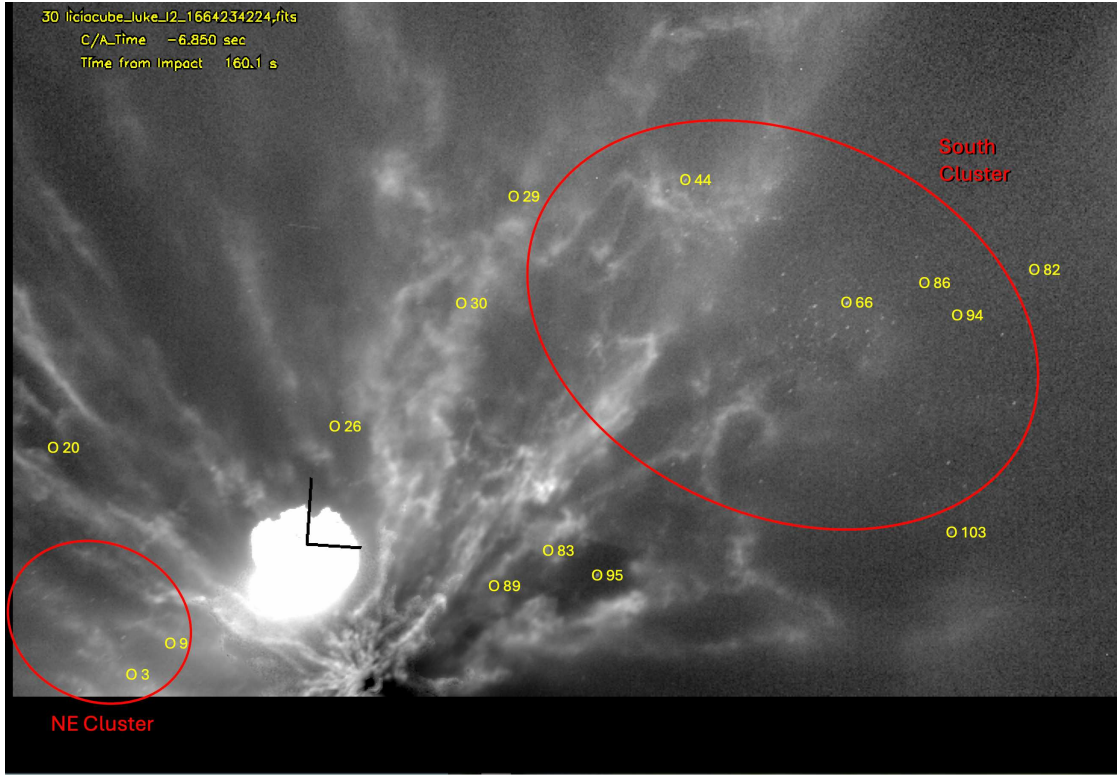

Figure 6. Animated sequence showing clusters of boulders to the upper right and lower left, as well as a general lack of boulders in other directions. The sequence covers the 12 seconds before LICIACube's close approach, showing the apparent motion of the boulders due to the changing viewpoint. Numbered circles highlight a selection of 16 boulders that are tracked through the sequence, and discussed in Section 5. Images were processed by dividing out a $r^{-1.6}$ radial profile.
[This figure links to an animated sequence]

As the flyby phase progresses, the ejecta continues turning away from LICIACube, and the material in the lower left quadrant again returns to the field of view. At ~178 sec, the far wall of the dust cone crosses behind Dimorphos (Figure 4e) and in subsequent images, the full circle of ejecta again becomes visible (Figure 4f). Although this is essentially an inverted view of the inbound leg, the projected angles of the cone are substantially different, providing a unique perspective of the scene. Only a few boulders are detected in the post-C/A images, glimpsed as they pass through gaps in the streamers (discussed in more detail in Section 6.)

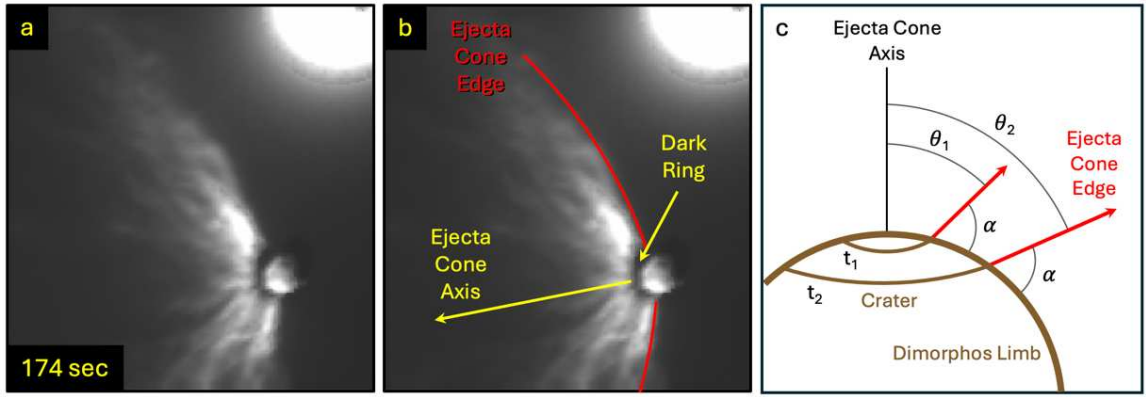

Figure 7. a) Image showing the walls of the ejecta cone, seen from nearly perpendicular to its axis, and b) the same image with annotations highlighting the curvature of the walls. c) diagram illustrating how the curvature of Dimorphos' surface causes the angle between the cone's axis and the outflowing ejecta ($\theta$) to increase with time as the crater grows, while the angle with respect to the local surface ($\alpha$) remains constant. Panel b) also highlights the dark ring at the base of the ejecta cone, which is caused by optically thick dust that blocks sunlight from penetrating through to the back of the cone.

An interesting feature that becomes visible in the post-C/A images is a dark ring at the base of the dust cone adjacent to Dimorphos, extending out to >100 m (Figure 7b) (this feature was also highlighted by Deshapriya et al. (2023), Dotto et al. (2024), Cheng et al. (2024) and Zinzi et al. (2024)). On first inspection, it is not clear if this ring represents a gap where the ejecta has detached from Dimorphos (implying a cutoff in the ejection), or if it is a shadow on the ejecta itself. If it is a gap, then it must have opened up in the few seconds between the last approach image and the first departure image of Dimorphos, because no gap was observed during the approach. This rapid opening would require ejection speeds of tens of meters per second, which is inconsistent with the observations, suggesting it is not a gap. Furthermore, as the sequence progresses, the bright wall of the cone in the background can be seen moving from the left to the right, passing behind Dimorphos at ~178 sec, yet the ring remains dark throughout this time period. If the ring was a gap, the bright background ejecta would be visible through the opening. Thus, we conclude that the dark ring is not a gap, but instead, is the result of a dense cloud of ejecta above the impact site casting a shadow on the material beneath. This leads to the conclusion that, at 170 sec after the impact, the innermost ~100 m of the dust cone is dense enough to be optically thick and blocks sunlight from passing through to the backside. This result confirms that the shadow seen on Dimorphos during the approach observations is indeed being cast from the dense inner regions of the ejecta cloud.

In the post-C/A images, Dimorphos itself presents a striking figure. It is clear that, three minutes after impact, the body remains a solid ellipsoid and thus survived the impact process. It also exhibits numerous characteristics that inform about the body itself as well as providing indirect information about the ejecta. These characteristics are repeated in numerous frames with different exposure times (Figure 8), indicating that they are real features and not artifacts introduced in the merging of the image triplets. First, the sunlit face of the surface can be seen, while the night side is silhouetted against the background ejecta, providing a full-disk view of the body. Comparing the illumination of Dimorphos to Didymos reveals that there is a shadow, cast by the optically thick ejecta at the base of the cone, across the upper limb of Dimorphos, which is likely to be an extension of the shadow seen in the approach images. As the sequence progresses, the shadow appears to narrow. This behavior is expected, because it is disappearing over the limb as the sub-spacecraft point moves from -40° latitude toward the equator. The shadowed region is slightly brighter than the night side of the asteroid, suggesting that either some sunlight is passing through the ejecta that is casting the shadow, or, more likely, the bright ejecta overhead is providing indirect illumination of this region, in the same way that the ejecta from the Deep Impact experiment illuminated terrain on the night side of Tempel 1 (Thomas et al. 2007). Finally, Dimorphos exhibits bright structures that extend beyond the expected terminator of a triaxial ellipsoid (night-side topography). These features indicate one of several possibilities: 1) There are topographical features (e.g., boulders or slopes) that rise up high enough to be illuminated, even when the Sun is > 90° from the zenith; 2) the side of Dimorphos not imaged by DART does not conform to the oblate spheroid shape derived from the DART observations; or, less likely, 3) material on the side of Dimorphos opposite the DART impact was lifted due to seismic shaking.

Zinzi et al. (2024) performed an analysis of the night side of Dimorphos in these images to measure the ratio of its long and short axes. They suggest that the body may be more elongated than was measured by the DART observations and conclude that there is a shadow on the sunward limb in the post-C/A LUKE data. In our reprocessed images, we easily detect both the sunward limb (unshadowed) and the silhouetted night-side limb, allowing us to directly determine the extent of the body. Measurements from the images in Figure 7 give average dimensions of 175±6 m along the long axis and 118±5 m along the short axis (ratio of 0.67). These dimensions are consistent with those derived in the shape model by Daly et al. (2024) of 179±1 m and 115±1 m. (Because the sub-spacecraft latitude ranged from -40° to -26° in these observations, the short dimension in the image does not exactly correspond to the body's shortest axis.)

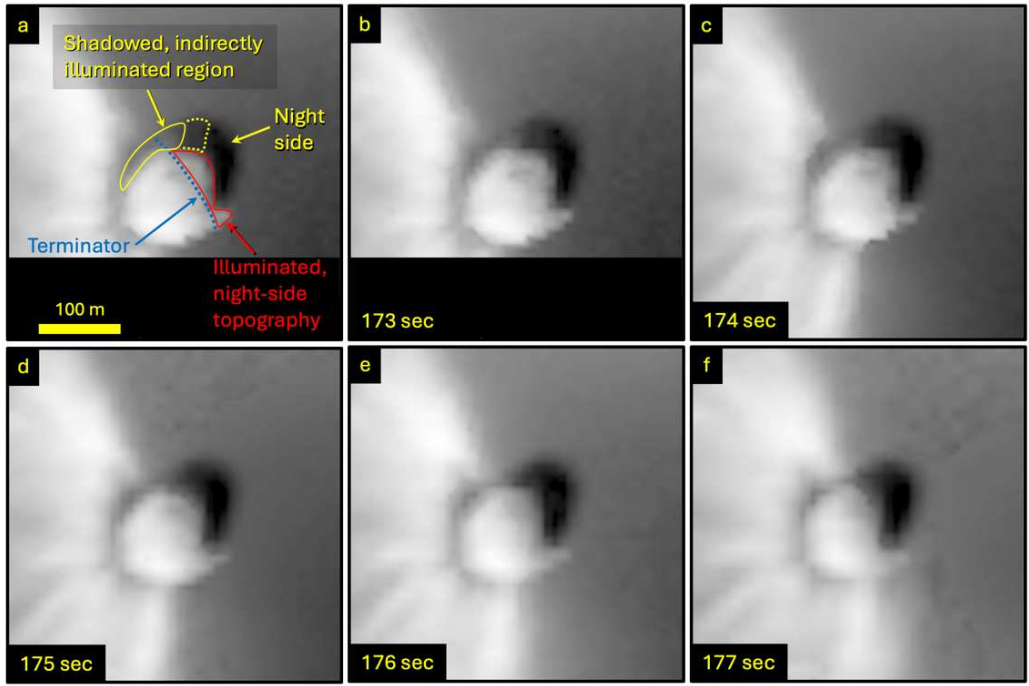

Figure 8. Temporal sequence of images with close-ups of Dimorphos after close approach. Features noted in the text are highlighted in panel a (173 sec), including the illuminated night side topography and the shadow along the top limb cast by the dense ejecta at the base of the cone. This shadowed region is somewhat brighter than the night side due to indirect illumination from the bright ejecta higher in the cone. (Note that this indirect illumination extends beyond the terminator.) Over time, the shadow also disappears over the horizon as the spacecraft moves from -40° to -26° latitude during the sequence. The images are displayed with a logarithmic display to more clearly reveal the illuminated night-side topography.

Because Didymos is not obstructed by the ejecta, it was imaged from dramatically different geometries over the course of the flyby (Figure 4). As noted earlier, many details are seen on the surface, and jagged edges along the terminator and concavities along the limb suggest it is not strictly triaxial.  Starting at ~172 sec and continuing throughout the rest of the observations, the dark limb of Didymos can be seen in silhouette against a diffuse uniform sky background (discussed in more detail in Section 3.4).  The limb definition seen in these silhouettes will provide critical constraints for modeling the full shape of Didymos.

It has been noted that there is significant glare off the sunward limb of Didymos (Figure 4d-f), and there is consideration as to whether this represents secondary dust, kicked off the surface of Didymos by ejecta from Dimorphos, or simply scattered light due to instrumental optics (Mazzotta Epifani et al. 2023).  The flyby sequence provides clues to help address this issue.  As noted above, the bright edge of the dust cone passes behind Dimorphos at ~178 sec.  At this time, the viewpoint lies parallel to the edge of the cone, with the bulk of the dust concentrated to the left of this edge.  There is a significant gap between the dust and Didymos (approximately 300 m from the closest limb), suggesting that the vast majority of the material missed the asteroid by hundreds of meters (Hirabayashi et al. 2024).  Although there is no evidence for it, we acknowledge the possibility that some of the DART ejecta (or possibly the vapor plume discussed in Section 3.4) reached Didymos and produced a small amount of secondary dust, but we suggest that the majority of the glare is due to scattered light from the surface of Didymos.  There are several lines of evidence supporting this conclusion.

First, the glare is very smooth, showing no structure, and remains consistently so regardless of changes in the viewing geometry.  It is not likely that secondary dust ejecta would be this uniform and would instead exhibit bright and dark regions (e.g., higher intensity on the side closer to Dimorphos or streamers originating where boulders landed).  Second, the spatial distribution of the glare around Didymos' limb correlates well to the brightness of Didymos' adjacent surface.  This means it peaks along the sunlit limb and stops at the terminator, even though the sunlit limb is roughly perpendicular to the incoming ejecta and the terminator is face-on.  This remains true as the LUKE viewpoint changes, causing different portions of the surface to lie at the limb, which again suggests that the glare is simply scattered light.  Finally, three images (1663814377, 166381452, and 1663814487) of a crescent Earth were obtained in the days before impact and these images exhibit a halo around the Earth.  Although the crescent is small, this glare is detectable out to 10-15 pixels from the limb.  In this case, the glare cannot be attributed to dust, so these observations show that the LUKE instrument has an uncorrected scattered light issue.  Didymos is much brighter and larger in extent than the crescent Earth and it would be expected that the scattered light would scale up proportionately. The existence of scattered light means that it is necessary to characterize how much of the glare is produced by scattered light before it can be determined how much of the glare around Didymos might come from secondary dust ejection.

### 3.3. Departure Phase

In comparison to the approach and flyby phases, the departure phase is relatively uneventful. This phase is defined by images from 195 to 243 sec, where radial motion again dominates over translational motion. It is a short segment and the evolutionary changes are not seen as clearly as in the approach phase because the fast ejecta has left the field of view, the evolution of the slower-moving morphology is offset by the decreasing resolution, and scattered sunlight rapidly causes the background to rise, overwhelming the fainter aspects of the debris field (the camera points within 63° of the Sun at the end of the sequence). However, these images can still be used to measure the velocities of clumps and kinks in the streamers. Furthermore the features may be linked to those seen during the approach, providing a different perspective for evaluating the structure and evolution of the debris field.

### 3.4 Post-Close Approach Sky Background / Plume

As noted in Section 3.2 and seen in Figure 4f, Didymos is seen in silhouette after C/A, indicating that LICIACube is viewing the asteroid with illuminated ejecta in the background. First distinguishable at 172 sec, Didymos' nightside limb remains faintly visible until ~183 sec, at which time the ejecta cone has rotated behind Didymos and the background level increases dramatically. This raises the question: What is the source of the faint background illumination in the ~10 sec before the ejecta cone rotates behind Didymos? At 172 sec, the viewpoint is ~70° from DART's trajectory and the nightside limb is far outside edges of the observed cone. Thus, in order to be located behind Didymos, the material providing the background illumination must have been emitted at very high speed and at a wide angle with respect to the DART trajectory. We suggest that the most likely explanation is the vapor plume.

The vapor plume was first reported by ground-based observations within hours of impact (it was the first visual evidence confirming the impact) and was also observed by the Lucy spacecraft (Graykowski et al. 2023; Fitzsimmons et al. 2023; Weaver et al. 2024). The physics of what produced this plume have not yet been fully explained, but it was observed as a hemispherical cloud expanding outward, with speeds up to ~2 km sec$^{-1}$, from the Didymos system in the general direction opposite of the DART trajectory. During LICIACube's approach, this plume was situated in front of Didymos and thus, would not have produced a silhouette. With the plume moving toward LICIACube, the spacecraft would have passed through it between ~110 and ~140 sec. As LICIACube tracked Didymos through the flyby, it was turning to look back at the plume, and apparently started to see its effect behind Didymos at 172 sec. We attempted to support this proposed scenario by measuring the sky brightness over time, looking for changes when the plume was expected to be in or out of the field of view, but were not successful. The sensitivity of the camera, imperfect calibration and reduction procedures, and contamination from the ejecta cone preclude any solid conclusions about variations in the sky brightness.

## 4. Boulder Measurements

As noted above, over 100 boulders were detected in multiple images. Because of LICIACube's motion during the flyby, the changing viewpoint introduced parallax in their apparent positions, and this information can be used to determine their absolute positions and velocities relative to Dimorphos. Furthermore, the photometric brightness can be used to estimate the boulder sizes. The results from these measurements then allow an analysis of the spatial distribution of the ejected objects, as well as a measure of the momentum that they carry.

### 4.1. Determining 3-D Boulder Locations and Velocities

The basic procedures and equations for deriving an object's position from two images are described in Appendix 1. Once a position from two images has been found, we then improve the solution by adding measurements from multiple images to reduce the statistical errors and to allow a determination of the object's linear velocity. Starting with the initial estimate from the parallax analysis, we include an initial guess of the velocity, and then perform a multi-variable fit to find the position and velocity vectors that best reproduce the locations of the boulder in each of the images where it is measured. (SPICE transformation matrices are used to convert from J2000 inertial positions to the pixel locations in the LUKE images.) The boulder's position vector is defined for the time of LICIACube's close approach and the velocity is assumed to be constant.

In deriving our solutions, we took two different approaches and compared the results. In the first, we perform a 3-variable fit to determine the position vector of the boulder, and then adopt the velocity vector as simply the position vector divided by 167 sec. This approach assumes that the boulders were ejected in an impulsive event at the time of impact and are travelling in linear trajectories. The impulse assumption is valid to within a few percent, because it is likely that the boulders were ejected within the first few seconds in order to impart the high velocities that are measured (Housen et al. 1983). Also, the clustering of the boulders is difficult to explain if they are ejected over a longer period of time. The linear trajectories are valid because even the slowest boulders in our sample have velocities that are well above the escape velocity of the Didymos system (0.25 m s$^{-1}$ in the vicinity of Dimorphos).

In our second approach, we perform a 6-parameter fit, where we solve for both the position and velocity vectors. Comparing to the previous result, the position vectors agree well (within ~3%), but the velocities in this approach show variability that depends on which, and how many, observations are included. This is due to the fact that the boulders are only moving a few pixels during the ~10 second observation window, so the pixel measurement uncertainties have a significant influence on the resulting velocity vectors. Even so, the velocity vectors in the two approaches agree to within 10-20%.

Although the 6-parameter fit produces slightly better residuals to the observed pixel positions, this is due to the mathematical effect of more free parameters, rather than being a better reflection of reality. Because the 6-parameter fits are derived from observations spanning only ~10 seconds, the directional components of the velocities have uncertainties that produce large positional errors when projected backward 167 seconds. Thus, the trajectory's origins do not converge on Dimorphos. For this reason, we have adopted the solutions from the 3-parameter fit to represent our results, to produce a self-consistent result that originates at the impact site.

### 4.2. Image Processing and Measurement Procedures

For our work, we require a high-quality image dataset and analysis techniques that allow us to accurately identify and track objects over multiple frames. We performed our own data reduction, as described in Section 2.2, to produce images with a sky background that permits the identification of boulders in as many images as possible. Only the red images were used, because they have a narrower PSF and better contrast than the other colors. To further improve our ability to detect and measure the boulder locations against the sky background, we made use of different scaling algorithms (e.g., linear or log) for displaying the images, as well as utilizing two different enhancement techniques to improve the visibility of features. Our first enhancement was the division of a $r^{-1.6}$ radial profile, centered at the impact site, which suppresses the bright central region of the dust cone so that detail can be seen at all distances in a single display. Our second enhancement is an unsharp mask, an edge-finding technique that works well to sharply highlight the boulders. We make it a point to use multiple displays to visually confirm that any features we are tracking are not being introduced by the image manipulations. Our measuring tool, discussed below, has the ability to switch between different enhancements and displays to allow these checks to be done in parallel with the measurement process. However, we always return to the original image as a reality check and use it to measure the feature's location.

To perform our measurements, we adapted a tool that was developed for use in our analyses of the large grains observed in the inner coma of comet 103P/Hartley 2 by the Deep Impact eXtended Investigation (DIXI) (Hermalyn et al. 2013). This interactive tool allows the user to select an object in one image and then step through other images to track the object, measuring its location at each step. (Measurements are relative to the center of Didymos and then offset to the impact site, as noted in Appendix 1). The identification of the boulder in the first two images is governed by the object's brightness and relative position with respect to other objects. After measurements have been obtained in two or more images, the object's position and velocity are computed on-the-fly as described in Section 4.1. The solution is then used to predict where the object would be expected to appear in other images. With just a few data points, the solution is quickly refined so that subsequent predictions tend to be very accurate. This gives us the ability to track a specific boulder, even within a crowded cluster of objects. A good solution is characterized by the measurement residuals typically matching the predicted positions to within a few pixels. This residual can be attributed to several sources: 1) uncertainty in the position of the reference point, Didymos; 2) uncertainties in the twist angle on the sky; and 3) uncertainties in the pixel position itself. The predictive capability also allows the boulders to be tracked long-term throughout the flyby, revealing when an object, measured in the pre-C/A images, should re-enter the field of view after C/A. Another strength of the on-the-fly solution is that it reveals misidentified points, because the residuals to the fit notably increase when a "bad" measurement is included, instilling confidence that the identifications are accurate.

We validated our procedures and measuring tool using Dimorphos as a test case. Even though Dimorphos' exact location is difficult to define when it is hidden behind ejecta in the pre-C/A images, there are enough measurements (30 pre- and post-C/A) to produce an excellent solution that agrees well with the known orbital position. Our solution found a distance 1.20 km from Didymos, which compares well to the known semimajor axis of 1.18 km, and the individual X, Y and Z positions match well to the results from SPICE calculations for the time of impact. We also derived a radial velocity of 0.5 m s-1 which is small enough to be consistent with the zero radial velocity between Didymos and Dimorphos. These results provide confidence that our techniques and measurements can be used to accurately map the boulders in the Didymos environment. Machine-readable tables containing the measured X and Y positions and photometric brightnesses of the 104 boulders, in each of the 33 images where they appear, can be accessed from the University of Maryland Digital Repository at https://doi.org/10.13016/ehyk-zkn9 (Farnham 2025c).

### 4.3. Measurement Uncertainties and Selection Effects

There are a number of potential sources of error in our measurements, which can be difficult to quantify. First, there may be errors introduced by geometric distortion and field curvature. The LICIACube Software Interface Specification (Zinzi & Della Corte 2023c) states that the distortions are < 0.1%. However, Didymos moves from the center of the image to the corner and back during the course of the flyby and with an 8° × 4° field of view, the corners are where any field curvature or distortions are likely to be most pronounced. Thus even issues that introduce shifts of only a few pixels will introduce errors in our measurements. Second, there are uncertainties in the geometric parameters computed from our trajectory solution (Appendix 2). Although our solution was the best result we could derive, we see oddities that suggest there are still some minor issues that we were not able to resolve. These seem to be related to the twist angle on the sky around the time of close approach. Because Dimorphos was off the field of view during this time, and we relied on the relative positions of Didymos and Dimorphos to compute the twist angle, our computed value may be off by a degree or so. Even small twist angle offsets can produce errors of several pixels in boulders at large distances. Third, our measurements can be affected by the large field of view, because the viewpoint is different by as much as 10° across the diagonal corners of the field. Although we attempt to correct for this factor in our geometric computations, there are still uncertainties that are unquantified, and these change as the apparent locations of the boulders move relative to Didymos.

We explored how different measurements in different regions of the image affect our positional solutions. In general, the normalized residuals for our calculations are typically 2-3 pixels, with slightly smaller values closer to Didymos, and somewhat larger values for the most distant measurements. An analysis of these variations suggests that the uncertainty in our positional precision is typically 2-4%, with objects with more measurements tending to produce better results. Our quoted velocities are simply the position divided by 167 sec, so the uncertainties on the velocities are comparable.

Because the asteroids moved into the lower left corner of the images around the time of closest approach, we are not able to see any boulders that might be located off much of the bottom or left side of the frame. Thus, there are regions in space where we cannot track any objects. It is important to understand where these regions are located, so they do not affect our interpretation of the results. If the data are insufficient, then we can't make any conclusions about a lack of detections, but an absence of objects where they should be visible means that there are no boulders present. To this end, we performed a test to define the regions where solutions are precluded: we created a volumetric grid extending out to 10 km from Dimorphos and tested each point within this grid to determine if it is visible in at least four of the images where other boulders were detected. The results from this test revealed an "exclusion zone", where we can make no conclusions. More importantly, we found that for the rest of the sky, we should be able to track any boulders that *are* present to distances of at least 5 km from Dimorphos.

### 4.4. Anomalous Images

In the execution of the work, we found problems with two of the merged images that should be noted. First, frame 1664234228 exhibits a systematic offset between the pixel location of the boulders and that predicted from our solution, with the observed positions consistently located farther along the track than predicted. The most likely explanation for this offset is an error in the observation time, which should probably be ~0.2 sec later than reported. Measurements from this image are not used in computing our solutions. A second image, 1664234231, is the last frame in which boulders are seen during the approach phase. Some of the objects, especially those near the edge of the frame, exhibit dramatically larger residuals when the measurements from this image are included. We suspect that this is the result of non-optimal registration of the image, because Dimorphos is out of the field of view, and Didymos is only partially visible. Thus, we only include measurements from this frame when they were consistent with the general solution.

### 4.5. Photometry and Boulder Sizes

Our final step is to measure the photometry of the boulder in each image. We tested aperture sizes and found that a 3-pixel radius aperture, with the sky background derived from a 4-pixel wide annulus, maximized the S/N in typical images. Even so, the brightnesses tend to exhibit significant scatter, as discussed below. The images are calibrated to units of radiance, so our photometry is converted to irradiance by dividing by $(7.8 \times 10^{-5})^2$ (steradians covered by a pixel). We then correct for the spacecraft range to normalize all measurements to the close approach distance (57.63 km) and use the HG function from Bowell et al. (1989), with G=0.20, to normalize to 0° phase angle. (Given the 45° to 96° range of solar phase angles, adopting a value of 0.15 would produce a difference of ~0.1 mag, or a 10% difference in brightness, which is a minor component of the uncertainty, as discussed below.)

The measurements of boulder's brightness can vary by as much as a factor of four for objects seen in many (> 15) measurements. This scatter arises from several sources: The object's signal and the sky background are less well characterized in shorter exposure images; The background varies as the boulders cross over different ejecta clouds; The signal recorded is affected by the positioning of the PSF in the Bayer filter (brightest if it is centered on the red pixel, fainter if it is on a blue or green pixel); and more physically, the boulders are not necessarily round (Robin et al. 2024) and may be tumbling rapidly due to the impact, producing temporal variations in their lightcurves. These error sources suggest that the brighter measurements represent higher quality data, so we favor providing them more weight in combining the data into an averaged irradiance. Several techniques were tested, with bisquare weighting[2] generally producing the best results. However, when many measurements were obtained, the bisquare weighting frequently rejected one or more of the brighter points, which unintentionally gave the fainter measurements more weight. To avoid this problem, we adapted our technique to apply the bisquare weighting to the brightest 6 measurements for combining the data (adopting 4 to 8 measurements produced only

---

[2] https://github.com/wlandsman/IDLAstro/blob/master/pro/biweight_mean.pro

~10% difference in the final results). Additional discussion of the brightness uncertainties is included in Section 5.

After computing the averaged irradiance, we derive the effective boulder radius, $R$, in meters, using the relation $R = (4\ I_{meas}/I_{Sun}\ r^2\ \Delta^2) / p$, where $I_{meas}$ is the normalized, average irradiance, $I_{Sun} = 1.5$ W m$^{-2}$ nm$^{-1}$, $r = 1.046$ AU, $\Delta = 57{,}630$ m at close approach, and the albedo $p = 0.17$ (Naidu et al. 2020).

## 5. Boulder Measurement Results

We measured the positions, velocities and brightnesses of a total of 104 individual boulders, which range in size from 0.2 to 3.6 m in radius. We included only objects measured in at least four images, but the vast majority of objects have 6 to 10 observations, and some were measured in up to 21. Although our detection limit is ~0.2 m, it can be difficult to track fainter objects because they often do not appear in the shorter exposures and thus may not be uniquely identified across multiple frame gaps. Because of this, we estimate that there are an additional 30-40 objects ≤ 0.5 m that are seen but are too faint or in regions that are too crowded to track. Table 1 gives a subsample of properties of 16 representative boulders (highlighted in Figure 6) including the biggest, fastest, slowest, and other random objects. A more extensive table that includes all 104 objects, with additional parameters that include the XYZ positions and velocities in different coordinate systems, (for both our 3- and 6-parameter fits) and radii computed from our photometric measurements, is provided in the supplementary, machine-readable Table 1. [Machine Readable Table 1 goes here.]

Results are summarized in Figure 9, which plots the azimuth and elevation angle at which the boulders were ejected (e.g., the location in the sky where they would be seen by an observer at the impact site). The points are color coded to show their velocity and distance from Dimorphos and the point sizes denote the computed radii of the boulders. The tinted pink background shows the exclusion zone where the data preclude any conclusions about a lack of objects.

The boulders are not uniformly distributed, but tend to be concentrated in two basic populations with a few scattered objects in between. The primary group, containing ~70% of the measured objects, forms a dense cluster to the south (azimuths from 150-200°) with ejection speeds of 30-50 m s$^{-1}$. (Most of the untracked objects also seem to be part of this dense cluster.) Intriguingly, these objects have trajectories that lie between 4° and 25° of the local surface, which means that they are situated outside the dust ejecta cone (shown by the dashed blue line). The second group, containing ~15% of the objects, is less dense than the primary group and is located to the northeast (azimuths from 30-60°). Objects in this cluster lie at higher elevation angles, 25-50° (inside the cone), and have velocities 10-20 m s$^{-1}$.

**Table 1**. Representative properties of a select subsample of boulders

| Boulder Number | $r_{Dim}$[a] (km) | $v_{Dim}$[a] (m s$^{-1}$) | Long.[b] (deg) | Lat.[b] (deg) | Azim.[c] (deg) | Elev.[c] (deg) | Radius (m) |
|---|---|---|---|---|---|---|---|
| 3 | 1.75 | 10.5 | 298.2 | 24.5 | 37.0 | 35.7 | 1.1 |
| 9 | 1.50 | 9.0 | 305.1 | 16.1 | 50.1 | 36.9 | 1.2 |
| 20 | 2.99 | 17.9 | 328.1 | 4.5 | 75.0 | 23.3 | 0.5 |
| 24 | 1.34 | 8.0 | 336.3 | -48.1 | 133.9 | 26.1 | 3.6 |
| 26 | 2.30 | 13.8 | 339.4 | -49.9 | 135.8 | 24.1 | 0.6 |
| 29 | 4.78 | 28.6 | 355.0 | -68.7 | 156.4 | 17.4 | 0.7 |
| 30 | 3.87 | 23.2 | 344.6 | -69.1 | 156.6 | 21.2 | 0.6 |
| 44 | 5.92 | 35.5 | 15.3 | -80.3 | 169.4 | 15.3 | 0.7 |
| 66 | 5.89 | 35.3 | 99.1 | -78.2 | 181.7 | 7.5 | 1.2 |
| 82 | 7.36 | 44.1 | 113.8 | -74.5 | 186.3 | 5.3 | 0.7 |
| 83 | 2.69 | 16.1 | 157.1 | -82.4 | 186.4 | 16.4 | 1.3 |
| 86 | 7.74 | 46.4 | 160.2 | -82.0 | 186.9 | 16.7 | 0.6 |
| 89 | 2.19 | 13.1 | 191.3 | -81.8 | 187.2 | 21.1 | 1.1 |
| 94 | 8.80 | 52.7 | 188.5 | -78.6 | 190.7 | 21.2 | 0.4 |
| 95 | 2.52 | 15.1 | 123.7 | -70.7 | 190.8 | 3.6 | 1.9 |
| 103 | 6.70 | 40.2 | 161.0 | -68.6 | 200.0 | 12.7 | 0.8 |
| 104 | 1.06 | 6.36 | 196.2 | -51.7 | 219.5 | 28.1 | 1.8 |

[a] Distance and velocity with respect to Dimorphos
[b] Sub-boulder longitude and latitude with respect to Dimorphos coordinates
[c] Azimuth and elevation angle of the boulder as seen from the impact site on Dimorphos

Although we have no information about any objects that might be in the exclusion zone, we can state that otherwise only a few boulders exist outside of the two groups, and those are widely distributed in azimuth, with velocities 5-20 m s$^{-1}$. There are three boulders at ~190° azimuth (yellow and orange) that are much closer to Dimorphos than the others in that region, and so it is not clear if they are part of the primary, high-speed cluster or if they should be classified with the randomly distributed objects. The large number of objects within each of the two groups and the fact that they are clustered in space and have similar velocities, compared to the otherwise sparsity of boulders distributed in other directions, indicates there must be a connection between the objects in each group. e.g., they have the same origin where special conditions existed during the impact. This connection is discussed further in Section 6.1.

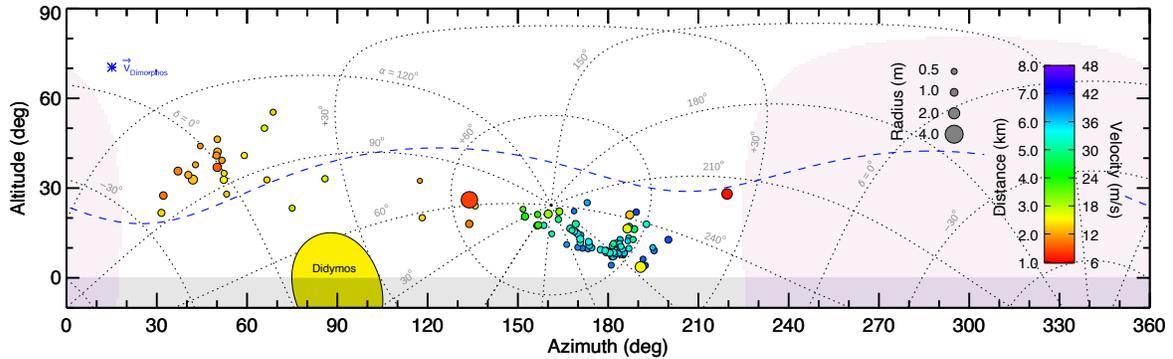

Figure 9. Plot showing the azimuth/elevation angle trajectories derived for the 104 boulders measured in this study (showing their locations in the sky if viewed from the impact site). Most of the objects are clustered in two groups, with only a few others scattered in between. Those in the larger cluster to the South were ejected on trajectories only 4° to 25° above the horizon. The points are sized to denote the boulders' radii and color coded to show their distances from Dimorphos. (Closer boulders are plotted on top, so in crowded areas, more distant objects may be hidden by closer ones.) The pink tinted region denotes the exclusion zone, where any boulders that might be present would not be seen because they were outside of the images when the spacecraft was close enough to detect them. The yellow ellipsoid represents the projection of Didymos, and the shaded bar at the bottom shows the region below the horizon. The dashed blue line represents the projection of the walls of the dust ejecta cone. Equatorial coordinates for the celestial sphere are plotted for reference. The asterisk at upper left indicates the direction of Dimorphos' velocity (linear momentum) vector, the direction defining the effective β value.

Most of the boulder measurements come from images from the approach phase. This is because the geometry and camera orientation are such that most of the boulders, even those at large distances, are projected onto the sky within the field of view until around C/A, when they all move off the left side of the image. After C/A, the more distant boulders remain outside the field until the spacecraft is too far away to detect them again, while most of the closer ones are hidden behind the dust cone. Only four objects are measured in both the pre and post-C/A images, all of which are larger in size and relatively close to Dimorphos.

The cumulative size distribution of our measured objects is plotted in Figure 10, which also provides a depiction of the uncertainties in the boulder radii (see Table 1). The average error on the boulder radius is ~65%. This uncertainty is dominated by the scatter in the brightness measurements, which typically vary by a factor of 1.5-2.5 with a few extreme cases that exhibit variations as large as a factor 4. However, when the measurements are plotted as temporal lightcurves for each boulder, they usually exhibit trends of increasing or decreasing brightness with time rather than random scatter, which suggests that the variations are real changes that reflect the rotation of irregularly shaped objects. Thus, the accuracy of our measurements is probably better than the scatter would indicate. Unfortunately, because of the lightcurve variations, we can't quantify this improvement, but it is likely good enough to reduce the uncertainties on the radius to better than 50%. This then propagates into uncertainties on the density and momentum of factors of 2-3.

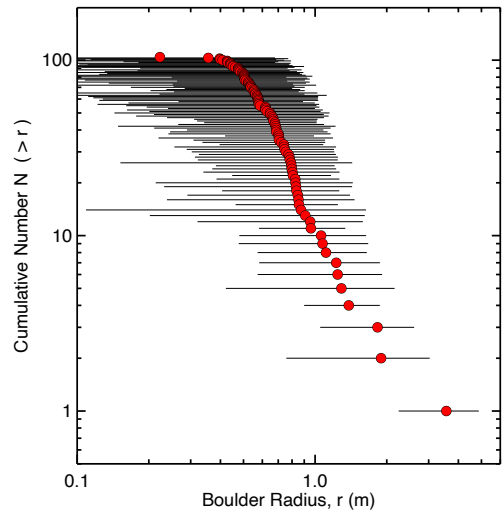

Figure 10. Plot showing the cumulative size distribution of the 104 ejected boulders measured in this study, with error bars that reflect the general uncertainties on the radii.

A simple power law slope could be fit to the boulder size distribution, but this may not be meaningful for several reasons: 1) The uncertainties in the measurements are large; 2) The measurements represent potentially three separate families of objects; and 3) Several dozen other objects in the same size range are observed but not tracked (and thus would not be represented in the size distribution). Such a fit could be informative, but it would require a detailed analyses to address these caveats, which is beyond the scope of this paper.

## 6. Discussion
### 6.1. Boulder Origins and Trajectories

Measurements of the boulder locations and trajectories obtained in this work have revealed a number of interesting constraints on the early impact processes. Specifically, there are a number of observables that need to be explained:

1) Numerous boulders were detected, with radii as large as 3.6 m.
2) The boulders are not spatially correlated to the asymmetric dust cone, with the vast majority located either inside or outside the ejecta cone.
3) The boulders are primarily grouped in two clusters. One cluster is concentrated to the south, is spatially dense, and its constituents were ejected at high velocities (30-50 m s$^{-1}$) and low angles to the surface (outside the dust cone). A second cluster to the northeast (NE) is less dense, has lower velocities (<20 m s$^{-1}$), and higher ejection angles (inside the dust cone).
4) There is a demonstrable lack of boulders outside of these clusters, with only a few sparse objects in between. These isolated boulders include the largest objects and have velocities similar to those in the NE cluster.

Unfortunately our trajectory solutions are not accurate enough to pinpoint the exact locations on the surface where the boulders originated, and as such, they could have come from anywhere in the immediate vicinity of the impact site. However, the details of the impact surface allow us to infer the potential origins of the ejected boulders. More detailed modeling and simulations will be required to fully understand the implications of these observations.

First, because the boulders are not spatially correlated with the ejecta cone, they are likely the result of unique ejection mechanisms. Second, the lack of boulders in between the clusters suggests that they were not randomly distributed around the impact point; otherwise comparable numbers of boulders would be seen in all directions. Third, the high velocity boulders indicate that they came from the immediate proximity of the impact (because velocities drop rapidly with distance (Housen et al. 1983; Hirabayashi et al. 2024)) and were probably expelled by direct interactions with the spacecraft itself. Figure 11 shows the impact site, with the spacecraft bus and solar panels outlined (Daly et al. 2023). Two large boulders closely straddle the impact site, Atabaque to the South and Bodhran to the North. Each of these boulders was struck by a solar panel, probably shattering the boulders into meter-sized fragments, a fraction of a second before the spacecraft bus hit the surface. Given the correspondence in direction, it is highly likely that the cluster of boulders ejected to the south are the remnants of Atabaque. This is consistent with laboratory experiments by Cline et al. (2023), which showed that boulders on the surface in the immediate vicinity of an impact can be ejected at high velocities and shallow angles with respect to the surface. The other random boulders ejected to the south may also be related to the destruction of Atabaque, but were ejected more peripherally, and at lower speeds. (The exact pattern of ejecta will depend on the specific solar panel geometry, surface normals, and interactions between surrounding boulder fragments.)

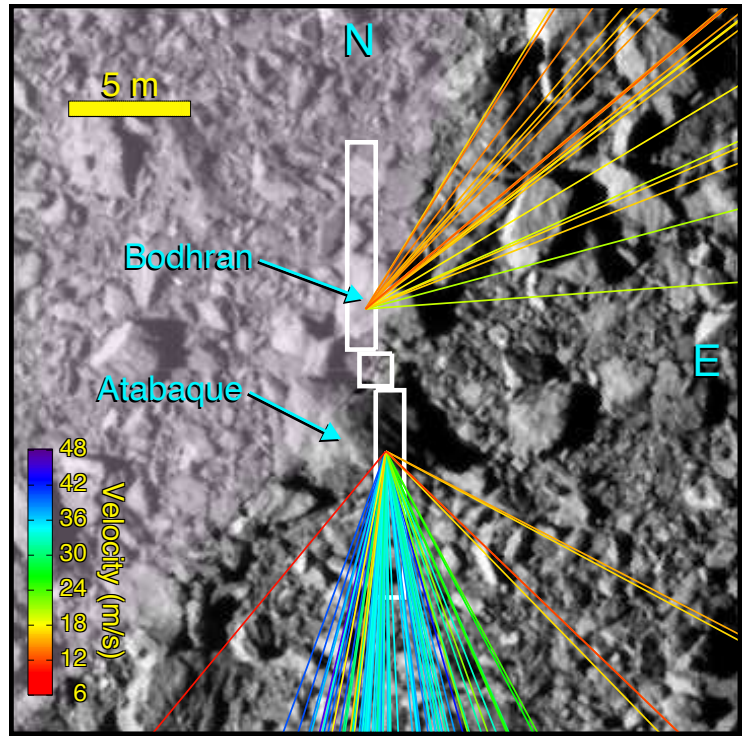

Figure 11. DRACO image of the impact site (0401930048_45552), with the spacecraft and solar panels outlined (white). Azimuthal trajectories of the 104 measured boulders are overplotted, with colors (reflecting the velocity) that match those in Figure 9. Although there are no constraints on their actual points of origin, a likely scenario is shown, with the southern cluster expanding radially from the boulder Atabaque and the NE cluster arising tangentially from Bodhran. It is possible that additional boulders are expanding radially from Bodhran, but these would be located in the exclusion zone (pink shading) where they would not be detectable in LICIACube observations. See text for additional discussion.

Connections between Bodhran and the ejected boulders are not as clear. It is logical that the events in this direction would be analogous to those involving Atabaque, in which case a dense, high-speed cluster of boulders would have been ejected radially Northward (in the direction of Bodhran). Unfortunately, if boulders were ejected in this direction, they would be located in the exclusion zone where they would not be detected in LICIACube observations (see Figure 9). However, in this scenario, the boulders to the NE could be analogous to the lower velocity bodies that are seen in the periphery around the southern cluster and thus would be related to the destruction of Bodhran. The differences in character between the two groups could be due to the details of the solar panel orientations: the -Y solar panel impacted Atabaque edge-on, while the +Y panel hit Bodhran face-on (Ernst et al. 2023).

If we assume objects are spherical, then the total volume of all 104 boulders amounts to 403 m$^3$, which is equivalent to a sphere 4.6 m in radius. Adopting the bulk density of Dimorphos, 2400 kg m$^{-3}$ (Daly et al. 2023), suggests that the mass ejected in these objects is $9.7\times10^5$ kg (~0.02% of Dimorphos' mass). This is a lower limit, because the density of the monolithic boulders is likely higher than the bulk density of the rubble pile Dimorphos, because there are additional boulders that we detected but could not track, and because there are likely more boulders in the exclusion zones that were not detected. The combined volume of the cluster of 74 high-speed southern boulders is ~90 m$^3$. For comparison, the volume of Atabaque (radius 3.3 m; Daly et al. 2023) is 150 m$^3$, which is sufficient to encompass all 74 of the boulders, plus the additional boulders that were observed but not tracked, and presumably many smaller bodies (down to pulverized dust) that must also exist.

All of the boulders measured in this work are moving fast enough to have escaped from the Didymos system and their ejection velocities put them into independent heliocentric orbits that differ slightly from Didymos'. To explore these orbits, we used the boulders' state vectors (heliocentric position at the time of impact and the boulder ejection velocity added to the heliocentric velocity of Didymos) to derive Keplerian elements for each object (Moulton 1970). We find that the ejection speeds are small compared to the heliocentric orbital velocity, so the boulder orbits exhibit only small changes relative to Didymos. For example, the most extreme change in perihelion distance changed from Didymos' value of 1.01295 AU to 1.01273 AU (0.02% change). Thus, none of the observed boulders will approach the Earth due solely to their ejection in the impact event. Future work that includes other forces is necessary to determine if the boulder orbits may evolve into Earth- or Mars-crossing objects on long timescales (e.g., Fenucci & Carbognani 2024; Peña-Asensio et al. 2024).

## 6.2. Boulder Energy and Momentum

The primary goal of the DART mission was to measure the momentum enhancement in the direction of Dimorphos' velocity vector, to prove the feasibility of using a kinetic impactor to change an asteroid's orbit. Results from the experiment showed a value $\beta = 3.6^{+0.19}_{-0.25}$ (Cheng et al. 2023), where β=1 is equivalent to the spacecraft momentum (579 kg × 6145 m s$^{-1}$ = 3.6×10$^6$ kg m s$^{-1}$). However, it is clear that significantly more momentum is contained in the ejecta (e.g., about half the momentum in the ejecta cone is directed sideways), but because it did not directly affect the orbital velocity, it has not been accounted for. Furthermore, the ejecta cone is not symmetric (Deshapriya et al. 2023; Hirabayashi et al. 2024) and is not aligned with Dimorphos' velocity vector (~20° offset), so these additional components of momentum could have potentially significant influences on Dimorphos' orbit and rotation state.

Because we have a measure of both the size and velocity of the boulders in this study, we can estimate their contributions to both the impact kinetic energy and momentum budgets. Summing contributions for the 104 boulders, we find they carry a total kinetic energy 1.6×10$^8$ J, which represents only about 1.4% of the energy of the DART spacecraft. We also find the boulders contain a total momentum 1.1×10$^7$ kg m s$^{-1}$. This is distributed in the Dimorphos body reference frame as [2.8×10$^6$, -1.2×10$^6$, -1.1×10$^7$] kg m s$^{-1}$ in the X, Y and Z directions, which in units of β, is [0.78, -0.33, -3.06]. The 0.33 component in the Y direction is aligned with the Dimorphos' velocity vector, which indicates that these boulders contributed ~9% of the momentum that changed the orbital period. Given the uncertainties on the boulder sizes (the largest unknown in the momentum calculations) discussed in Section 5, the uncertainty on the momentum results is probably within a factor of 2–3.

Of equal interest, however, is how the components in the X and Z directions affect Dimorphos' orbit and rotation state. With 96% of the momentum from the boulders directed to the South, it is also likely that Dimorphos' orbital plane inclination was changed by the impact, and if we assume that the pre-impact orbit was circular and coplanar with Didymos' equator, then we can estimate that change. From the remote observations, the orbital period was the only measurable quantity and this indicated a change of -2.7 mm s$^{-1}$ in Dimorphos' along-track velocity (Cheng et al. 2023; Naidu et al. 2024). We then use the momentum in the X and Z directions, with a bulk density of 2400 kg m$^{-3}$ and a volume equivalent radius of 75 m (Daly et al. 2024 2023; Richardson et al. 2024), to compute $\Delta v$ in the radial and normal directions of 0.7 and 2.6 mm s$^{-1}$, respectively. (We note that a radial $\Delta v$ of 0.7 mm s$^{-1}$ is within the uncertainty range given in the orbital solution of Naidu et al. (2024)). Assuming the binary orbit is Keplerian, which is sufficient for estimating the change in the inclination, we find that the changes in velocity in the along-track, radial, and normal directions would excite Dimorphos' inclination to ~0.85°. This corresponds to vertical excursions above and below Didymos' equatorial plane of nearly 20 m, which, while not measurable with ground-based observations, should be easily measured by the Hera mission when it arrives at the Didymos system (Küppers et al. 2022).

Although there are a lot of uncertainties and unknowns in these calculations – including whether there were boulders from Bodhran ejected to the North that would cancel out some of the southern momentum – the detection of any change in inclination has significant implications. Dimorphos is thought to have formed from the accretion of material shed from Didymos' equator (e.g., Walsh, Richardson, and Michel 2008; Agrusa et al. 2024), so its pre-impact inclination should be near zero[3]. Therefore, any orbital inclination measured by Hera should be the result of the DART impact. Dimorphos' inclination should be the best-preserved orbital element by the time of the Hera arrival, as tidal dissipation damps the inclination much more slowly than the semimajor axis and eccentricity. The time evolution of the semimajor axis and eccentricity can be approximated as (Goldreich & Sari 2009),

$$\frac{da}{dt} = 3 \frac{k_p}{Q_p} \frac{M_s}{M_p} \left(\frac{R_p}{a}\right)^5 na \qquad [1]$$

---

[3] Although Dimorphos is thought to have formed on a near-zero inclination orbit, it is possible that the binary orbit has been perturbed since formation, via impacts or planetary encounters. However, both the impact and planetary encounter timescales for the Didymos system are longer than the inclination damping timescale derived in this work (Richardson et al. 2022). In addition, all pre-impact observations were consistent with Dimorphos being on a zero inclination orbit, similar to population of other similar binary systems (See Richardson et al. 2024 and references therein).

$$\frac{de}{dt} = en\left[\frac{57}{8}\frac{k_p}{Q_p}\frac{M_s}{M_p}\left(\frac{R_p}{a}\right)^5 - \frac{21}{2}\frac{k_s}{Q_s}\frac{M_p}{M_s}\left(\frac{R_s}{a}\right)^5\right], \qquad [2]$$

where $k$ is the tidal Love number, $Q$ is the tidal quality factor, $M$ is the mass, $R$ is the body radius, $n = \sqrt{GM_p/a^3}$ is the mean motion, and the subscripts p and s stand for the values of the primary and secondary, respectively. The damping of inclination is a much slower process, and in the limit of low $e$ and $i$, can be approximated as (Burns 1977),

$$\frac{di}{dt} \approx -\frac{i}{4a}\frac{da}{dt}. \qquad [3]$$

The values of $Q/k$ are poorly constrained for rubble pile asteroids, so for simplicity we assume $\frac{Q_p}{k_p} = \frac{Q_s}{k_s} \approx 10^5$, although estimates of $Q/k$ vary by orders of magnitude in the literature, typically between $10^4$ and $10^7$ (e.g., Goldreich and Sari 2009; Jacobson and Scheeres 2011; Nimmo and Matsuyama 2019). Then, based on the nominal body masses, radii, and post-impact semimajor axis, we find the immediate post-impact rate-of-change of $a$, $e$, and $i$, are $\frac{da}{dt} \approx 0.004$ m yr$^{-1}$, $\frac{de}{dt} \approx -2.2 \times 10^{-6}$ yr$^{-1}$, and $\frac{di}{dt} \approx -7.4 \times 10^{-7}$ deg yr$^{-1}$, corresponding to respective eccentricity and inclination damping timescales of $1.3 \times 10^4$ yr and $1.1 \times 10^6$ yr. Although these numbers are highly uncertain because the values of $Q/k$ are poorly constrained and this calculation neglects various higher-order effects such as the BYORP effect and the non-spherical shapes of both bodies. However, this makes it clear that the inclination damping timescale is orders of magnitude longer than eccentricity damping. Therefore, even in cases with extreme tidal damping, it can be reasonably expected that the inclination will not meaningfully change before Hera's arrival.

Momentum imparted by the boulder ejection is also likely to have altered the rotational state of Dimorphos. The momentum in the X direction (parallel to the equator) produced a torque that acted to slow down the body's rotation, which will contribute to changes in the asteroid's spin rate. In addition, the momentum in the -Z direction produced a massive torque, primarily around the X axis, which would have driven Dimorphos into a tumbling state with potentially significant out-of-plane rotations. In fact, given that Dimorphos is a slow rotator that, in a single impulse, experienced a drastic change in its moment of inertia and massive torques perpendicular to its rotation axis, it seems inconceivable that it did not enter a tumbling state as a result of the impact (cf., Agrusa et al. 2021; Richardson et al. 2024). The damping timescales of tumbling systems are difficult to estimate, due to their chaotic nature (e.g., Meyer et al. 2023), but they tend to be longer than the five years that mark the arrival of Hera. Thus, it is likely that Hera will observe Dimorphos in a state of non-principal axis rotation.

### 6.3. Comparison to Boulders seen in HST Data

Jewitt et al. (2023) observed 37 boulders surrounding Didymos in Hubble Space Telescope (HST) images obtained on 19 December 2022 and 4 February 2023. These objects range from 0.75 to 3.75 m in radius, with most ~2 m in radius. Although these are in the same size range of our boulders, the fact that they exhibit velocities of only ~0.5 m s$^{-1}$ (projected on the sky), means that this is a completely different population of boulders. The low speeds suggest that they were excavated and ejected at a later time in the event. Given the projected velocities, the HST boulders would still remain within 100 m of Dimorphos at the time of the LICIACube observations and thus would be embedded in the dense cloud of ejecta seen in our observations, where they would not be visible.

Although true velocities were not measured for these objects, we can estimate their momentum contribution. Jewitt et al. compute a mass of $5\times10^6$ kg and a median projected velocity of 0.3 m s$^{-1}$, which give a rough estimate of the total momentum ~$1.5\times10^6$ kg m s$^{-1}$. Because they were ejected late in the excavation process, their spatial distribution probably coincides with that of the ejecta cone in general, so the direction of their momentum contributions also reflect that of the ejecta cone. Given the opening angle of the cone, this means that about half their momentum is in the direction of the velocity vector, suggesting that these boulders may contribute ~0.2 of the $\beta$=3.6 experiment result.

Jewitt et al. (2023) also suggested that seismic shaking could be an alternate ejection mechanism for these objects, with the shock of impact propagating through Dimorphos and expelling the boulders from the surface opposite the impact site. LICIACube images from ~174 sec after impact clearly show the limb of the asteroid opposite the impact, and there is no evidence for any material being ejected at the speeds the boulders are traveling. (Section 3.2 notes possible evidence for seismic shaking to explain features at the surface of Dimorphos, but this material would be moving at sub-escape speeds.) The HST boulders must be traveling at speeds >0.25 m s$^{-1}$ to escape the Didymos system, and would thus be >40 m from Didymos, so unless they are in front of the illuminated surface or behind the disk, we would easily detect meter-sized boulders in these images. Thus, this evidence suggests that seismic shaking does not appear to be highly efficient on a rubble pile object like Dimorphos.

## 7. Conclusions

We provided a tour of the LICIACube flyby of the post-impact DART debris field, highlighting some of the interesting phenomena that were observed. This overview provides a guide for potential future studies of Didymos, Dimorphos and the ejected material. One of the most surprising features was the observation of scores of boulders ejected in specific directions. Our 3-D mapping of these objects indicated that most of them were ejected at shallow angles to the surface, with high velocities, and they seem to be the remnants of large boulders shattered in the impact. These results, along with the filamentary structures in the ejecta cone, will provide unique insights into the physics regarding impacts into rubble pile asteroids, where material on and under the surface can have important implications on the temporal and spatial evolution of the ejecta (e.g., Arakawa et al. 2020; Kadono et al. 2020; Kadono et al. 2022; Ormö et al. 2023; DeCoster et al. 2024). These boulders also represent significant momentum contributions that were not accounted for in the orbital period measurements (due to the direction they were ejected), suggesting that accounting for the momentum of the ejecta in all directions could reveal many times the $\beta=3.6$ value that was measured opposite Dimorphos' velocity vector in the DART experiment. Thus, a full accounting of the momentum in all directions and understanding the role played by surface boulders will provide better knowledge of how the specifics of the impact could alter – either reducing or enhancing – the effects of a kinetic impactor. Finally, the contributions from the boulder momentum suggest that Hera may encounter a Didymos system in which Dimorphos is tumbling in an orbit that has a slight inclination to Didymos' equator.

## APPENDIX 1. 3-D Positions From Parallax

We use measurements of the boulder's position in two images obtained at different times to derive the location of a boulder in space. For convenience, we define a close approach reference frame (CARF), centered on Didymos. In this frame, the +X axis is parallel to the LICIACube velocity vector, and the +Z axis is the vector from Didymos to the point of LICIACube's closest approach (see Figure 12). Using SPICE matrix transformations, we convert the boulder's apparent pixel position, projected onto the plane of the sky in each image, to the CARF ($\mathbf{P}_1 = [X_{P1}, Y_{P1}, Z_{P1}]$ and $\mathbf{P}_2 = [X_{P2}, Y_{P2}, Z_{P2}]$), and also compute the position of the spacecraft ($\mathbf{S}_1 = [X_{S1}, Y_{S1}, Z_{S1}]$ and $\mathbf{S}_2 = [X_{S2}, Y_{S2}, Z_{S2}]$) for the same times (See Appendix 2).

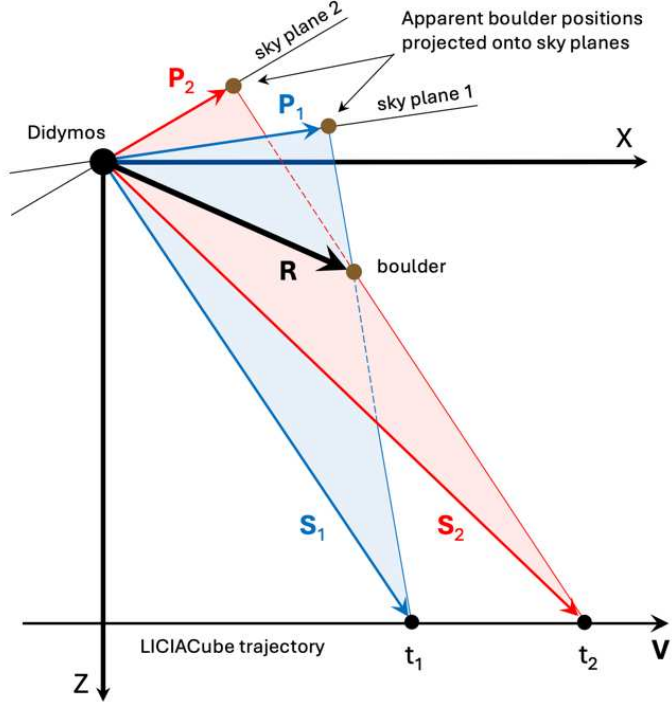

Figure 12. Illustration showing how the parallax between two LUKE images obtained at different times ($t_1$ and $t_2$) is used to derive the 3-dimensional position of a boulder. For each observation, three points (LICIACube's position, Didymos, and the apparent position of the boulder on the sky plane) define a plane. The intersection of the two planes defines the position of the boulder in three dimensions. The close approach reference frame defined in the text is used for convenience.

For each observation, the spacecraft, Didymos and the projected position of the boulder form a plane. The line of intersection of the two planes under consideration is the vector, $\mathbf{R}$, which defines the boulder's position with respect to Didymos. To determine the direction of $\mathbf{R}$, we use the cross product to find the normal vectors for each plane, $\mathbf{N}_1 = \mathbf{S}_1 \times \mathbf{P}_1$ and $\mathbf{N}_2 = \mathbf{S}_2 \times \mathbf{P}_2$ and then derive the unit vector that defines line of intersection from $(\mathbf{N}_2 / |\mathbf{N}_2|) \times (\mathbf{N}_1 / |\mathbf{N}_1|) = \hat{\mathbf{R}} = [i_R, j_R, k_R]$.

We then obtain the boulder's position vector $\mathbf{R} = [X_R, Y_R, Z_R]$ using the requirement that two points along a line satisfy the requirement:

$$\frac{X_R - X_{S1}}{X_{P1} - X_{S1}} = \frac{Y_R - Y_{S1}}{Y_{P1} - Y_{S1}} = \frac{Z_R - Z_{S1}}{Z_{P1} - Z_{S1}} \qquad [A1].$$

Since **R** is colinear with its unit vector,

$$\frac{X_R}{i_R} = \frac{Y_R}{j_R} = \frac{Z_R}{k_R} \qquad [A2]$$

and

$$Y_R = X_R \frac{j_R}{i_R} \quad \text{and} \quad Z_R = X_R \frac{k_R}{i_R}. \qquad [A3]$$

Separating $X_R$ from Eq. A1 and replacing $Y_R$ from Eq. A3,

$$X_R = X_{S1} + \left(X_R \frac{j_R}{i_R} - Y_{S1}\right)\left(\frac{X_{P1}-X_{S1}}{Y_{P1}-Y_{S1}}\right) \qquad [A4]$$

then solving for $X_R$ gives

$$X_R = \left[X_{S1} - Y_{S1}\left(\frac{X_{P1}-X_{S1}}{Y_{P1}-Y_{S1}}\right)\right]\left[1 - \left(\frac{j_R}{i_R}\right)\left(\frac{X_{P1}-X_{S1}}{Y_{P1}-Y_{S1}}\right)\right]^{-1}. \qquad [A5]$$

Solving for $Y_R$ and $Z_R$ using Eq. A3 defines the position of the boulder relative to Didymos. (As a check, we perform the same computation using the line of sight from image 2, and confirm that it gives the same result.)

Finally, we use SPICE matrix transformations to convert the CARF coordinates into the J2000 frame, and then apply an offset from the center of Didymos to the impact site on Dimorphos, ([0.8959, 0.7290, 0.2970] km) to translate the boulder positions so they are relative to the impact site at the time of closest approach. We can then convert to other coordinate systems (e.g., impact site azimuth and elevation angle) as needed using additional SPICE transformations.

APPENDIX 2.   LICIACube Trajectory Solution

Accurate geometric information was unavailable when this work was underway, so we computed our own spacecraft trajectory and viewing geometry using data contained in the images. This effort was possible because some of the images contain background stars that constrain the LICIACube pointing, while the relative positions of Didymos and Dimorphos define the orientation of that pointing on the sky (e.g., the direction of Celestial North). We found five different stars that appear in the background of 14 different images throughout the flyby, and used a preliminary trajectory to identify each star. We then measured the offsets in RA and Dec between the star and Didymos, which establishes the relative direction of the spacecraft when each image was obtained.[4] With this information, we were then able to solve for a trajectory that reproduces the spacecraft locations of those 14 images. Table A1 lists the pixel positions of Didymos and the star in each image, as well as the position angle of the North Celestial Pole that was needed to match the positional offsets for that image. Because the result is a continuous path, we can then use it to compute the geometric conditions for the LICIACube at any time between impact and the end of the LUKE observations.

---

[4] To define the position of Didymos, we adopted the SPICE kernels defined in the DART metakernel dart_v03.tm (https://naif.jpl.nasa.gov/pub/naif/pds/pds4/dart/dart_spice/).

Table A1. Star measurements used in the LICIACube trajectory solution

| LUKE Image (16642+) | $t_{imp}$ (s) | Didymos Pixel $X_D$ | $Y_D$ | $PA_{NCP}$[a] (deg) | Star Pixel $X_s$ | $Y_s$ | Star Coords.[b] $RA_s$ | $Dec_s$ | Star Name |
|---|---|---|---|---|---|---|---|---|---|
| 34099_00051 | 35.11 | 939 | 394 | 41.3 | 1397 | 639 | 310.01 | -18.14 | υ Cap |
| 34111_00051 | 47.09 | 937 | 396 | 41.2 | 1489 | 639 | 310.01 | -18.14 | υ Cap |
| 34117_01025 | 53.12 | 926 | 400 | 41.1 | 1530 | 643 | 310.01 | -18.14 | υ Cap |
| 34123_00076 | 59.10 | 925 | 401 | 41.0 | 1590 | 643 | 310.01 | -18.14 | υ Cap |
| 34135_00051 | 71.09 | 925 | 404 | 40.9 | 1731 | 645 | 310.01 | -18.14 | υ Cap |
| 34159_00051 | 95.09 | 930 | 401 | 40.7 | 89 | 385 | 316.78 | -25.01 | 24 Cap |
| 34171_00051 | 107.09 | 929 | 401 | 40.6 | 426 | 382 | 316.78 | -25.01 | 24 Cap |
| 34195_00051 | 131.09 | 929 | 381 | 38.8 | 1737 | 351 | 316.78 | -25.01 | 24 Cap |
| 34225_00112 | 161.07 | 330 | 129 | 10.2 | 1307 | 519 | 6.57 | -42.31 | α Phe |
| 34237_00112 | 173.07 | 276 | 205 | 330.2 | 945 | 760 | 82.06 | -20.76 | β Lep |
| 34256_00245 | 192.08 | 726 | 554 | 325.8 | 56 | 565 | 114.83 | 5.23 | α CMi |
| 34259_00523 | 195.09 | 753 | 579 | 326.0 | 531 | 531 | 114.83 | 5.23 | α CMi |
| 34264_00062 | 200.07 | 798 | 572 | 325.5 | 1157 | 448 | 114.83 | 5.23 | α CMi |
| 34271_00079 | 207.07 | 831 | 595 | 325.6 | 1781 | 393 | 114.83 | 5.23 | α CMi |

[a] Position angle of the North Celestial Pole (twist angle)
[b] Celestial coordinates from the Bright Star Catalog (Hoffleit & Warren 1995)

To define LICIACube's trajectory, we need to determine its velocity vector, **V**, and the time, $t_{CA}$, and location, **r**, of close approach relative to Didymos (Figure 13). To this end, we first assume LICIACube was on a linear trajectory with a speed 6.144 km s$^{-1}$ (adopting DART's speed, which is sufficiently accurate for the ~300 sec of the path we are considering). The point of close approach lies in a plane perpendicular to the velocity vector and passing through Didymos. This plane is derived by computing the vector $\mathbf{E} = \mathbf{P} \times \mathbf{V}$, where **P** is the rotation axis of Didymos. Because **E** is perpendicular to both **P** and **V**, it lies in both the C/A plane and Didymos' equatorial plane. We then rotate **E** around **V** by an angle $\alpha$, to produce vector **r**, which defines the direction of LICIACube when it is at C/A. Adopting this geometry, we can solve for the optimum trajectory using five variables: The direction (RA and Dec) of vector **V**, the close approach distance (e.g., magnitude of vector **r**), the rotation angle, $\alpha$, and the time of close approach, $t_{CA}$ (where $t=0$ is the time of the DART impact).

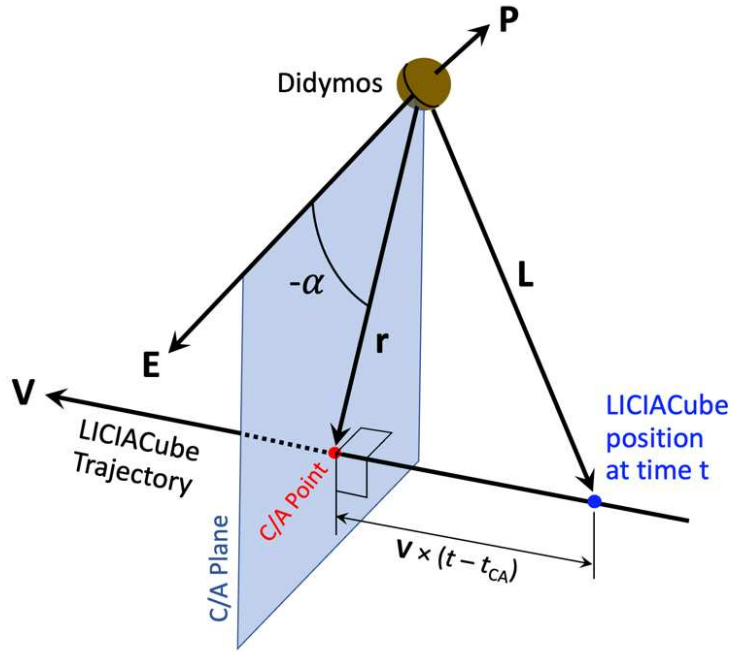

Figure 13. Illustration of the geometry used for deriving the LICIACube trajectory. The point of close approach (**r**) occurs in a plane that is perpendicular to the velocity vector **V** and passes through Didymos. Note that the angle $\alpha$ is measured in the direction of the right hand rule, but is shown for a negative value to match the geometry of this trajectory. See text for additional discussion.

We performed a five-variable, least-squares fit to the trajectory, constrained by the star measurements from the 14 images, and found a solution that reproduces the expected star positions to better than 0.05°. With this trajectory, we computed the pointing geometry of the LUKE camera throughout the flyby, and used measurements of the clock angle of Dimorphos relative to Didymos to determine the direction of celestial North in each image. In the instances where Dimorphos was out of the field, we had no measurements of the clock angle, so we assumed that both the clock angle and the rate of change of the clock angle changed smoothly with time, in an attempt to reproduce the acceleration and deceleration of the spacecraft slewing across the gap. These new constraints on the celestial North direction provide a better measure of the offsets in RA and Dec between Didymos and the stars, and we used them to perform a second iteration to improve our final trajectory. The apparent predicted path of Didymos, with comparisons of the measured and predicted star positions, is shown in Figure 14.

Our final result produces a velocity vector in the direction of RA, Dec = [307.52°±0.18°,-17.61°±0.18°], a close approach distance of 57.63±0.82 km, a rotation angle -56.97°±0.65°, and the time of close approach of 166.94±0.17 sec. In vector notation (J2000 coordinates), **V**= [3.567, -4.645, -1.859] km s$^{-1}$ and **r** = [-25.438, -34.830, 38.223] km. With the velocity and close approach vectors known, the location of the spacecraft can be determined using $\mathbf{L} = \mathbf{r} + \mathbf{V}(t - t_{CA})$. Using this new trajectory and pointing, we generated SPK and CK SPICE kernels for computing the geometry of the LUKE images[5]. These kernels will be archived with the Planetary Data System.

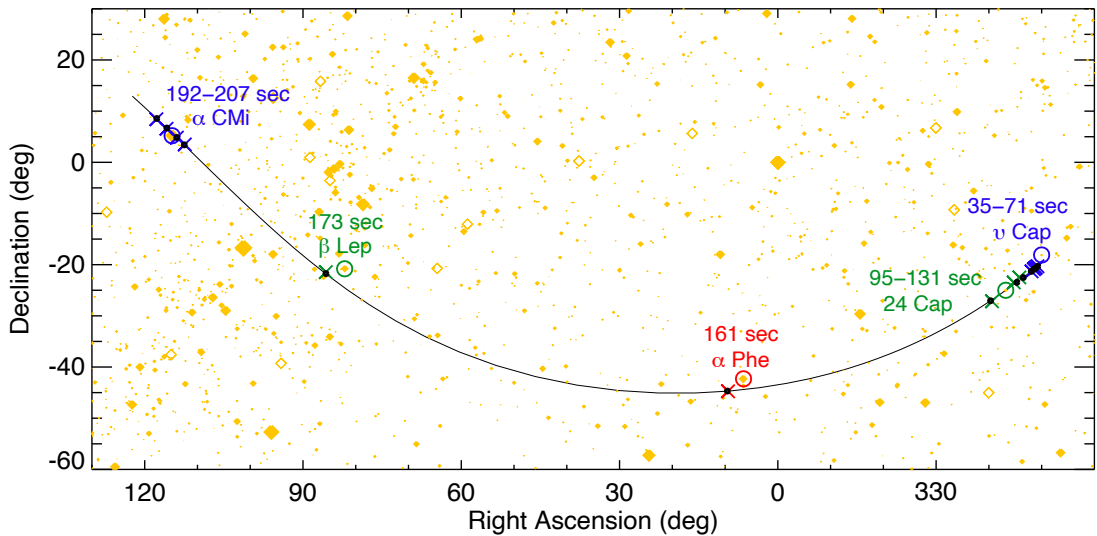

Figure 14. Plot showing the apparent path of Didymos across the celestial sphere (thin black line), as computed from the final LICIACube trajectory. The five stars used to derive the trajectory are highlighted by the colored circles, with the positions of Didymos measured in the 14 corresponding images shown by crosses of the same color. The small black dots denote the positions of Didymos predictions for the times of each image, with their correspondence to the crosses indicating the good fit of the trajectory solution. Yellow dots are stars from the Bright Star Catalog, with the size reflecting their brightness.

In our work with the new trajectory we computed the relative apparent position of Dimorphos with respect to Didymos. We found that the results matched well to the images before close approach, but that after close approach, Dimorphos was farther from Didymos than predicted, by as much as 4 pixels (assuming that Dimorphos' position during the flyby is at its pre-impact semi-major axis distance). This offset suggests that there is a small error in the pre-impact orbital solution for the Didymos system (JPL 542). The flyby images are particularly sensitive to the relative positions of the two bodies, so we were able to explore how altering the semimajor axis and the pole orientation of Dimorphos' orbit could improve the prediction of its position relative to Didymos. Using a three variable, least squares fit to the measured separation distance, we find that the best pole is oriented toward RA, Dec = [70.3°, -74.8°] and the semimajor axis is 1.175 km. These results differ from the JPL Solution 542 by less than 1-σ (Pole RA, Dec = [69.33°, -72.72°], $a$ = 1.184 km), but they significantly improve the predicted position of Dimorphos after close approach. This information has been provided to the JPL navigators for inclusion in future derivations of the pre-impact orbital parameters. Until a new, comprehensive solution is produced, we adopt JPL solution 542 for this work, and accept that this will introduce a few pixels of systematic positional error for Dimorphos.

Acknowledgements

We would like to thank Kosuke Kurosawa and an anonymous reviewer for helpful comments that improved the manuscript. This study was supported in part by the DART mission, National Aeronautics and Space Administration (NASA) contract No. 80MSFC20D0004 to JHU/APL. H.A. was supported by the French government, through the UCA J.E.D.I. Investments in the Future project managed by the National Research Agency (ANR) with the reference number ANR-15-IDEX-01. Portions of this work were performed under the auspices of the U.S. Department of Energy by Lawrence Livermore National Laboratory under Contract DE-AC52-07NA27344. LLNL-JRNL-2002297. R. L. and D. G. acknowledge support by the NASA/GSFC

---

[5] Named licia_2022_269_2022_269_tlf_v01.bsp and licia_2022_269_2022_269_tlf_v01.bc

Internal Scientist Funding Model (ISFM) Exospheres, Ionospheres, Magnetospheres Modeling (EIMM) team, and the NASA Solar System Exploration Research Virtual Institute (SSERVI). R. L. and D. A. acknowledge work done through the Center for Research and Exploration in Space Science and Technology (CRESST-II) supported by NASA award number 80GSFC24M0006. The LICIACube team members acknowledge financial support from Agenzia Spaziale Italiana (ASI, contract No. 2019-31-HH.0 CUP F84I190012600).